\documentclass[12pt]{article}

\usepackage{adjustbox}
\usepackage{algorithm}
\usepackage{algorithmicx}
\usepackage{algpseudocode}
\usepackage{amsmath}
\usepackage{amssymb}
\usepackage{amsthm}
\usepackage{array}
\usepackage[english]{babel}
\usepackage[backend=biber,
			style = nature,
			citestyle=numeric,
			sorting=none]{biblatex}
\usepackage{booktabs}
\usepackage{caption}
\usepackage{comment}
\usepackage{fancyhdr}
\usepackage{fourier}
\usepackage{fullpage}
\usepackage{graphicx}
\usepackage{hyperref}
\usepackage[utf8]{inputenc}
\usepackage{multirow}
\usepackage{pgfplots}
\usepackage{pifont}
\usepackage{rotating}
\usepackage{soul}
\usepackage{subfig}
\usepackage{tabularx}
\usepackage[most]{tcolorbox}
\usepackage{threeparttable}
\usepackage{tikz}
\usetikzlibrary{arrows}
\usetikzlibrary{shapes.geometric,calc}
\usepackage{times}
\usepackage[colorinlistoftodos]{todonotes}
\usepackage{tabularx}
\usepackage[normalem]{ulem}
\usepackage{url}
\usepackage{wrapfig}
\usepackage{xcolor}

\usepackage{rotating,lipsum}
\usepackage{etoolbox}

\useunder{\uline}{\ul}{}
\hypersetup{
    colorlinks=true,
    linkcolor=blue,
    filecolor=magenta,      
    urlcolor=cyan,
    pdftitle={Overleaf Example},
    pdfpagemode=FullScreen,
    }
\newcommand{\beginsupplement}{%
        \setcounter{table}{0}
        \renewcommand{\thetable}{S\arabic{table}}%
        \setcounter{figure}{0}
        \renewcommand{\thefigure}{S\arabic{figure}}%
        }
     
\newcommand{\thickhline}{%
    \noalign {\ifnum 0=`}\fi \hrule height 1pt
    \futurelet \reserved@a \@xhline
}     

\newcommand{\cmark}{\ding{51}}%
\newcommand{\xmark}{\ding{55}}%

\usepackage{makecell}

\newcolumntype{s}{>{\hsize=0.3\hsize}X}
\newcolumntype{m}{>{\hsize=0.5\hsize}X}
\newcolumntype{b}{>{\hsize=0.65\hsize}X}
\newcolumntype{B}{>{\hsize=1.3\hsize}X}

\usepackage[a4paper,left=0.8in,right=0.8in,top=1in,bottom=1in,footskip=.25in]{geometry}
\usepackage{adjustbox}

\usepackage[most]{tcolorbox}

\DeclareCaptionFont{blue}{\color{}}

\newcommand\wordcount{
    \immediate\write18{texcount -sub=section \jobname.tex  | grep "Section" | sed -e 's/+.*//' | sed -n \thesection p > 'count.txt'}
(\input{count.txt}words)}

\title{Methods for Large-scale Single Mediator Hypothesis Testing: Possible Choices and Comparisons\vspace{-0.5cm} }
\author{ Jiacong Du$^1$, Xiang Zhou$^1$, Wei Hao$^1$, Yongmei Liu$^2$, \\ Jennifer A. Smith$^3$ and Bhramar Mukherjee$^1$\\
{\normalsize $^1$ Department of Biostatistics, University of Michigan, Ann Arbor, MI} \\
{\normalsize $^2$ Department of Medicine, Divisions of Cardiology and Neurology,} \\
{\normalsize Duke University Medical Center, Durham, NC } \\
{\normalsize $^3$ Department of Epidemiology, University of Michigan, Ann Arbor, MI }
}
\date{}

\begin{document}

\linespread{1}

\maketitle

\begin{abstract}
    Mediation hypothesis testing for a large number of mediators is challenging due to the composite structure of the null hypothesis, $H_0:\alpha \beta=0$ ($\alpha$: effect of the exposure on the mediator after adjusting for confounders; $\beta$: effect of the mediator on the outcome after adjusting for exposure and confounders). In this paper, we reviewed three classes of methods for multiple mediation hypothesis testing. In addition to these existing methods, we developed the Sobel-comp method, which uses a corrected mixture reference distribution for Sobel's test statistic. We performed extensive simulation studies to compare all six methods in terms of the false positive rates under the null hypothesis and the true positive rates under the alternative hypothesis. We found that the class of methods which uses a mixture reference distribution could best maintain the false positive rates at the nominal level under the null hypothesis and had the greatest true positive rates under the alternative hypothesis. We applied all methods to study the mediation mechanism of DNA methylation sites in the pathway from adult socioeconomic status to glycated hemoglobin level using data from the Multi-Ethnic Study of Atherosclerosis (MESA). We also provide guidelines for choosing the optimal mediation hypothesis testing method in practice. (word count: 196) 
\end{abstract}


\newpage 

\section{Introduction}

Mediation analysis is often used to identify potential mechanistic pathways of the effect of an exposure on an outcome. It becomes increasingly popular in recent decades in epidemiology \cite{vanderweele2015explanation,pierce2014mediation,huang2015igwas,yang2017identifying,chen2020tobacco}. With the advances of high-throughput technologies in genomics studies, mediation analysis often requires analyzing a large number of potential mediators \cite{zhang2016estimating,zeng2021statistical}. These agnostic explorations of high-dimensional mediators allow researchers to investigate molecular traits associated with complex diseases that may be a result of socioeconomic inequalities, environmental pollution, or other exogenous factors. In particular, molecular epidemiological research has frequently considered the mediating role of DNA methylation (DNAm), and mounting studies have identified methylation differences at CpG sites as important mediators for various diseases such as cancer \cite{kulis2010dna,vanderweele2012genetic,wu2018mediation}, cardiovascular disease \cite{richardson2017mendelian} and diabetes \cite{grant2017longitudinal}.

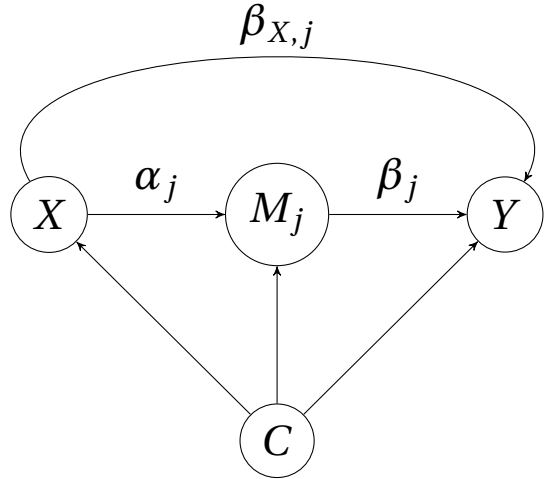
\begin{figure}
    \centering

\begin{tikzpicture}[->,>=stealth',auto,node distance=3cm,main node/.style={circle,draw,font=\sffamily\Large}]
\node[main node] (1) {$X$};
\node[main node] (2) [right of=1] {$M_j$};
\node[main node] (3) [right of=2] {$Y$};
\node[main node] (4) [below of=2]  {$C$};
\node at (3,2.5) {\Large $\beta_{X,j}$};

\draw [->] (1) -- (2) node [pos=0.5,above, font=\Large] {$\alpha_j$};
\draw [->] (2) -- (3) node [pos=0.5,above, font=\Large] {$\beta_j$};
\draw [->] (1) to [out=120,in=60] (3);
\draw [->] (4) -- (1);
\draw [->] (4) -- (2);
\draw [->] (4) -- (3);
\end{tikzpicture}
    
    \caption{A causal diagram for mediation analysis. For $j=1,2,...,J$, $X$ is the exposure, $M_j$ is the j-th mediator, $Y$ is the outcome, $C$ is the set of confounders. $\alpha_j$ is the effect of $X$ on $M_j$ after adjusting for $C$. $\beta_j$ is the effect of $M_j$ on $Y$ after adjusting for ($X,C$). $\beta_{X,j}$ is the direct effect of $X$ on $Y$ after adjusting for $M_j$ and $C$.}
    \label{illustration}
\end{figure}

Suppose there is a total number of $J$ mediators potentially mediating the effect of an exposure $X$ on the outcome $Y$. Let $M_j$ denote the j-th mediator where $j \in \{1,2,...,J\}$. To identify which $M_j$'s are truly in the mediating pathways, one can jointly model $M_1,M_2,...,M_J$ \cite{song2020bayesian, chen2018high, huang2019variance}. However, the computational burden may be too great and the solution may not be robust for large $J$ but modest sample sizes. Therefore, people often use the traditional univariate mediation analysis which examines one mediator at a time. This is often performed based on the parametric models proposed by \citeauthor{baron1986moderator} \cite{baron1986moderator}. For $j \in \{1,2,...,J \}$, the two regression models involved in a mediation analysis with continuous outcome and continuous mediators are: 
\begin{equation} \label{YMX_model}
    Y = \beta_{0,j} + \beta_{X,j} X +\beta_j M_j + \boldsymbol{\beta}_{C,j}^{\top} \boldsymbol{C} + \epsilon_{Y,j} 
\end{equation}
\begin{equation} \label{MX_model}
    M_j = \alpha_{0,j} + \alpha_j X + \boldsymbol{\alpha}_{C,j}^\top \boldsymbol{C} + \epsilon_{M,j}
\end{equation}

\noindent where $\boldsymbol{C}$ is the set of potential confounders and $\epsilon_{Y,j} \sim N(0, \sigma_{Y,j}^2)$ and $\epsilon_{M,j} \sim N(0, \sigma_{M,j}^2)$ are independent. The counterfactual framework, also called the potential outcome framework \cite{ vanderweele2015explanation,rubin1978bayesian}, is often used to define a causal mediation effect with certain accompanying assumptions \cite{valeri2013mediation, richiardi2013mediation,valente2017confounding}, including 1) no unmeasured confounders for the exposure-outcome relationship conditional on $\boldsymbol{C}$; 2) no unmeasured confounders for the mediator-outcome relationship conditional on $(X,\boldsymbol{C})$; 3) no unmeasured confounders for the exposure-mediator relationship conditional on $\boldsymbol{C}$; 4) no mediator-outcome confounders affected by $X$. Under the counterfactual framework with assumptions 1-4, the effect of the exposure on the outcome is decomposed into \textit{direct effect} and \textit{indirect effect} (also called \textit{mediation effect}). In addition, if there is no exposure-mediator interaction affecting the outcome, the causal estimate of the indirect effect is the same as the classical product estimate proposed by \citeauthor{baron1986moderator} \cite{baron1986moderator}. A causal diagram for illustrating the role of the j-th mediator is presented in \textbf{Figure \ref{illustration}}.

To test whether $M_j$ is mediating the effect of $X$ on $Y$, the underlying null and alternative hypotheses can be stated as:
\begin{equation*}\label{comp_test}
    H_{0,j}: \alpha_j \beta_j  = 0 \ \ vs. \ \ H_{1,j}: \alpha_j\beta_j \neq 0, \ \ for j=1,2,...,J
\end{equation*}
Since $H_{0,1},...,H_{0,J}$ are tested in a similar manner, we drop the subscript $j$ for now. 

The first class of methods contains Sobel's test \cite{sobel1982asymptotic} and the MaxP test \cite{mackinnon2002comparison}. The null hypothesis involving the product of parameters is composite \cite{barfield2017testing} and consists of three cases, namely, 1) $H_{01}: \alpha=0,\beta \neq 0$; 2) $H_{10}: \alpha \neq 0, \beta =0$; and 3) $H_{00}: \alpha=\beta=0$. Since Sobel's test and the MaxP test do not consider this composite structure, they are conservative \cite{barfield2017testing,liu2020large}, especially in high-dimensional settings where the majority of mediators are likely to have no mediation effect. 

Many recent studies have developed univariate analysis methods to produce calibrated p-values that consider the composite null structure. \citeauthor{huang2019genome} proposed the joint significance test under the composite null hypothesis (JT-comp) that uses the product of two normally distributed variables as the test statistic \cite{huang2019genome}. \citeauthor{dai2020multiple} developed a procedure for high-dimensional mediation hypotheses testing (HDMT) which considered the correct reference distribution for the MaxP statistic \cite{dai2020multiple}. A common feature for these two methods is to weight the reference distribution under $H_{01}, H_{10}, H_{00}$ to form a mixture null distribution corresponding to the test statistic. We group these two methods into the second class.

The third class contains the Divide-Aggregate Composite-null Test (DACT) method proposed by \citeauthor{liu2020large}. In contrast to the second class which form a mixture reference distribution, this method constructs a composite test statistic using the three p-values obtained under $H_{01}, H_{10}$ and $H_{00}$ \cite{liu2020large}. 

However, no study has numerically compared the testing performance of the above-mentioned methods. It remains unclear how these methods would be affected by various factors with high-dimensional mediators, in particular, by the sample size, the proportion of $H_{01}, H_{10}, H_{00}, H_1$ being true, the variation of non-zero $\alpha$ and $\beta$ across $J$ tests, and the $R^2$ in the data generating models, i.e. models \eqref{YMX_model} and \eqref{MX_model}. Our contribution in this paper is twofold. First, in addition to the existing methods, we develop a new method, called Sobel-comp, which is a variant of HDMT. Sobel-comp uses a corrected mixture reference distribution for Sobel's test statistic utilizing the composite structure of the null. Second, we perform extensive simulation studies to compare all six methods in three classes in terms of false positive rates under the null hypothesis and true positive rates under the alternative hypothesis. 

This paper is organized as follows: In Section 2.1, we first describe the five existing mediation hypothesis testing methods, including Sobel's test, MaxP, JT-comp, HDMT, and DACT, and discuss their potential advantages and limitations. We then propose our new method, Sobel-comp. In Section 2.2, we describe the simulation setup to compare the testing performance of the six methods. In Section 2.3, we describe the analyzing procedure for studying the mediation mechanism of DNAm in the pathway from adult socioeconomic status (SES) to glycated hemoglobin (HbA1c) level using data from the Multi-Ethnic Study of Atherosclerosis (MESA). Numerical results are presented in Section 3. We summarize the key strengths and limitations of each method and provide recommendations for applying these methods in practical settings in Section 4.

\section{Methods and Materials}

\subsection{Methods for mediation hypothesis testing}

Mediation hypothesis testing methods are often based on the Wald test statistics obtained from models \eqref{YMX_model} and \eqref{MX_model}. Denote $Z_\beta$ and $Z_\alpha$ as the test statistics for testing $\beta=0$ in model \eqref{YMX_model} and for testing $\alpha=0$ in model \eqref{MX_model}, respectively. Under the null hypothesis, we have:
$$Z_\alpha = \frac{\widehat{\alpha}-\alpha}{\widehat{\sigma}_{\alpha}} \sim N(0,1); Z_\beta = \frac{\widehat{\beta}-\beta}{\widehat{\sigma}_{\beta}} \sim N(0,1)$$
\noindent where $\widehat{\alpha}$ and $\widehat{\beta}$ are the maximum likelihood estimates for $\alpha$ and $\beta$, respectively. $\widehat{\sigma}_{\alpha}$ and $\widehat{\sigma}_{\beta}$ are the estimated standard error of $\widehat{\alpha}$ and $\widehat{\beta}$, respectively. Let the two-sided p-value for $Z_\alpha$ be $p_\alpha$ and for $Z_\beta$ be $p_\beta$. 

\subsubsection{Sobel's test}

Sobel's test statistic \cite{sobel1982asymptotic} uses the first-order multivariate delta method to find the standard error of $\hat{\alpha} \hat{\beta}$. Since $\widehat{\alpha}$ and $\widehat{\beta}$ derived from models \eqref{YMX_model} and \eqref{MX_model} are independent \cite{sobel1982asymptotic,mackinnon1995simulation}, Sobel's test statistic is defined as:
\begin{equation}
    T_{sobel} =  \frac{\hat{\beta}\hat{\alpha}}{\sqrt{\widehat{\beta^2}\widehat{\sigma_{\alpha}^2}+\widehat{\alpha^2}\widehat{\sigma_{\beta}^2}}} = \frac{Z_\alpha}{\sqrt{1+(Z_\alpha/Z_\beta)^2}}
\end{equation}

$T_{Sobel}$ is compared to $N(0,1)$ to determine the p-value. The null distribution of $N(0,1)$ is asymptotically correct under $H_{01}$ and $H_{10}$, but is incorrect under $H_{00}$, since the multivariate delta method fails at $(\alpha,\beta) = (0,0)$. \citeauthor{liu2020large} proved that $T_{Sobel}$ under $H_{00}$ asymptotically follows $N(0,1/4)$ \cite{liu2020large}. As a result, Sobel's test, which incorrectly uses $N(0,1)$ under $H_{00}$ as the reference distribution, yields larger p-values than the truth, and thus is conservative.

\subsubsection{MaxP test}

The MaxP test, also called the joint significance test \cite{mackinnon2002comparison}, has been developed based on the idea that if we want to reject $H_{0}$ at level $t$, we should reject two separate hypothesis tests of $\alpha=0$ and $\beta=0$ at level $t$ simultaneously. The MaxP test statistic is defined as:
\begin{equation}
    p_{max} = max \big(p_{\alpha},p_{\beta} \big)
\end{equation}
$p_{max}$ is compared to $U(0,1)$ to determine the p-value. Equivalently, $p_{max}$ is determined by the smaller $|Z_\alpha|$ or $|Z_\beta|$. Since $min(|Z_{\alpha}|, |Z_{\beta}|) > |T_{Sobel}|$ in a finite sample, the MaxP p-value is always smaller than that from Sobel's test and thus is more powerful. However, similar to Sobel's test, the reference distribution of $U(0,1)$ is incorrect under $H_{00}$. Since $P(p_{max}<t) = P(p_{\alpha}<t)\cdot P(p_\beta < t) = t^2$, the correct reference distribution for $p_{max}$ under $H_{00}$ is $Beta(2,1)$ \cite{liu2020large,dai2020multiple}. Since the p-value under $H_{00}$ determined by $U(0,1)$ will be larger than that by $Beta(2,1)$, the MaxP test is conservative.

\subsubsection{Joint significance test under the composite null hypothesis (JT-comp)}

We now resume to use the subscript $j$ corresponding to the $j$-th hypothesis test for $j=1,2,...,J$. The test statistic for JT-comp is the product of two normally distributed random variables, $Z_{\alpha,j} Z_{\beta,j}$ \cite{huang2019genome}. Unlike Sobel's test and the MaxP test, JT-comp 
distinguishes the null distributions for its test statistic under $H_{01,j},H_{10,j}$ and $H_{00,j}$ to obtain case-specific p-values. Specifically, let $w_{01,j},w_{10,j}, w_{00,j}$ be the probability of $H_{01,j}, H_{10,j}$ and $H_{00,j}$ being true, respectively. Denote $F(t)$ as the two-sided tail probability of the standard normal product distribution evaluated at $t$. Under $H_{00,j}$, since $Z_{\alpha,j} \sim N(0,1)$ and $Z_{\beta,j} \sim N(0,1)$, the case-specific p-value is $F(Z_{\alpha,j} Z_{\beta,j})$. Under $H_{01,j}$, $Z_{\alpha,j} \sim N(0,1)$ and $Z_{\beta,j} \sim N(\mu_{\beta,j}, 1)$, where $\mu_{\beta,j} = \beta_j/\widehat{\sigma}_{\beta,j} \neq 0$. \citeauthor{huang2019genome} further assumes that $\mu_{\beta,j}$ follows a symmetric distribution with mean $0$ and variance $\delta_{\beta,j}^2$, e.g. $\mu_{\beta,j} \sim N(0,\delta_{\beta,j}^2)$. By integrating out $\mu_{\beta,j}$, the p-value under $H_{01,j}$ is obtained by using the same $F(\cdot)$ function as if under $H_{00,j}$, but only differs by a scaling factor of $1/\sqrt{1+\delta_{\beta,j}^2}$. That is, the p-value under $H_{01,j}$ is $ F(Z_{\alpha,j}Z_{\beta,j}/\sqrt{1+\delta_{\beta,j}^2})$. Similarly, the p-value under $H_{10,j}$ is $F(Z_{\alpha,j}Z_{\beta,j}/\sqrt{1+\delta_{\alpha,j}^2})$, where $\delta_{\alpha,j}^2$ is the assumed variance of the mean of $Z_{\alpha,j}$ under $H_{10,j}$. The final composite p-value is aggregated as:
\begin{equation*}
    p_{JT-comp,j} = w_{01,j}F\Bigg(\frac{Z_{\alpha,j} Z_{\beta,j}}{\sqrt{1+\delta_{\beta,j}^2}} \Bigg) + w_{10,j}F\Bigg(\frac{Z_{\alpha,j} Z_{\beta,j}}{\sqrt{1+\delta_{\alpha,j}^2}} \Bigg) +
    w_{00,j}F\big(Z_{\alpha,j} Z_{\beta,j} \big)
\end{equation*}

$p_{JT-comp,j}$ is then approximated by Taylor series:
\begin{equation} \label{p_JTcomp_approx}
    \hat{p}_{JT-comp,j} = F\Bigg(\frac{Z_{\alpha,j} Z_{\beta,j}}{\sqrt{Var(Z_{\beta,j})}} \Bigg) +F\Bigg(\frac{Z_{\alpha,j} Z_{\beta,j}}{\sqrt{Var(Z_{\alpha,j})}} \Bigg) - F(Z_{\alpha,j} Z_{\beta,j})
\end{equation}

\noindent where $Var(Z_{\beta,j}) = 1+ w_{01,j}\delta_{\beta,j}^2$ and $Var(Z_{\alpha,j}) = 1+ w_{10,j}\delta_{\alpha,j}^2$. Sample variances of $Z_{\alpha,j}$ and $Z_{\beta,j}$ across all $J$ tests are used to estimate $Var(Z_{\alpha,j})$ and $Var(Z_{\beta,j})$. The advantage of using the approximation $\hat{p}_{JT-comp}$ is to avoid estimating $w_{01,j}, w_{10,j}, w_{00,j}$. Since the reference distribution of $Z_{\alpha,j} Z_{\beta,j}$ is correct under $H_{01,j}$, $H_{10,j}$ and $H_{00,j}$, JT-comp is more powerful than Sobel's and MaxP tests. 

However, the accuracy of $p_{JT-comp,j}$ approximated by $\hat{p}_{JT-comp,j}$ depends on the residual error from Taylor series expansion in \eqref{p_JTcomp_approx}. The residual error relative to the p-value becomes larger when the p-value becomes smaller \cite{huang2019genome}, suggesting that JT-comp cannot maintain the family-wise-error-rate (FWER) at small significance thresholds. A good approximation requires that $\delta_{\alpha,j}^2$ and $\delta_{\beta,j}^2$ are close to 0. 
Namely, the approximation works well when $\mu_{\alpha,j}$ is concentrated near zero ($\mu_{\beta,j}$ is similar). Since $\mu_{\alpha,j} = \alpha_j/\widehat{\sigma}_{\alpha,j}$, this condition is violated in cases such as having large $\alpha_j$; or having a large sample size so that $\widehat{\sigma}_{\alpha,j}$ is small. A practical suggestion given by \citeauthor{huang2019genome} is to check whether the sample variance of $Z_{\alpha,j}$ and $Z_{\beta,j}$ are less than 1.5. Since JT-comp only works well for small $\delta_{\alpha,j}^2$ and $\delta_{\beta,j}^2$, its applicability is limited to the settings with small samples and small $\alpha_j$'s and $\beta_j$'s.

\subsubsection{High dimensional mediation testing (HDMT)}

Another method which uses the correct reference distribution is HDMT \cite{dai2020multiple}. Let $\pi_{01},\pi_{10}, \pi_{00}$ be the proportion of $(\alpha_j=0,\beta_j\neq0), (\alpha_j \neq 0, \beta_j=0)$ and $(\alpha_j=\beta_j=0)$ among all $J$ tests. The test statistic for the HDMT method is the MaxP statistic. Under $H_{01,j}$ and $H_{10,j}$, $p_{max,j} \sim U(0,1)$ asymptotically. Under $H_{00,j}$, $p_{max,j} \sim Beta(2,1)$. The asymptotic reference distribution for $p_{max,j}$ is:
$$(\hat{\pi}_{01}+\hat{\pi}_{10})U(0,1) + \hat{\pi}_{00} Beta(2,1)$$
\noindent where $\hat{\pi}_{01}, \hat{\pi}_{10}$ and $\hat{\pi}_{00}$ are obtained by non-parametric methods for estimating the proportion of nulls\cite{storey2002direct}. It is worth mentioning that HDMT further proposes improving the power under finite samples. Under $H_{01,j}$, the p-value determined by $U(0,1)$ is accurate asymptotically when the power of rejecting $\beta_j=0$ goes to 1. Namely, $P(p_{\beta,j}<t|H_{01,j}) \xrightarrow{n\rightarrow\infty} 1$ for any $t>0$. However, this condition is difficult to hold when $t$ is extremely small in a finite sample, resulting in a noticeably larger p-value than the truth. In such cases, HDMT uses the Grenander estimator to estimate $P(p_{\beta,j}<t|H_{01,j})$ and $P(p_{\alpha,j}<t|H_{10,j})$. 

Overall, since the mixture null distribution of $p_{max,j}$ statistic is asymptotically correct, HDMT is robust to any choices of $\pi_{01},\pi_{10},\pi_{00}$. However, since the rejection rule of HDMT is determined by empirically estimating the significance thresholds and false discovery rates, it is difficult to compare it with other methods in terms of p-values. We make the following modifications to obtain p-values from HDMT using the asymptotic mixture reference distribution: 
$$p_{HDMT,j} = (\hat{\pi}_{01}+\hat{\pi}_{10})p_{max,j} + \hat{\pi}_{00}p_{max,j}^2$$

With finite samples, we estimate $P(p_{\alpha,j}<p_{max,j}|H_{10,j})$ and $P(p_{\beta,j}<p_{max,j}|H_{01,j})$ by the Grenander estimator as described in \cite{dai2020multiple}. The adjusted p-value is:
$$\Tilde{p}_{HDMT,j} = \hat{\pi}_{01} p_{max,j} \hat{P}(p_{\beta,j} < p_{max,j}|H_{01,j}) + \hat{\pi}_{10} p_{max,j} \hat{P}(p_{\alpha,j} < p_{max,j}|H_{10,j}) + \hat{\pi}_{00}p_{max,j}^2$$

\subsubsection{Divide-Aggregate Composite-null Test (DACT)}

The test statistic for DACT is constructed as a composite p-value obtained by averaging the three case-specific p-values weighted by $\pi_{01}, \pi_{10}, \pi_{00}$, respectively \cite{liu2020large}. Under $H_{01,j}$, the p-value is $p_{\alpha,j}$ since $\beta_j$ is known to be non-zero. Similarly, the p-value under $H_{10,j}$ is $p_{\beta,j}$. Under $H_{00,j}$, the p-value is $p_{max,j}^2$ using the MaxP statistic, which follows $Beta(2,1)$. The DACT test statistic is defined as:
\begin{equation} \label{DACT}
    DACT_j = \hat{\pi}_{01} p_{\alpha,j} + \hat{\pi}_{10} p_{\beta,j} + \hat{\pi}_{00} p_{max,j}^2
\end{equation}

\noindent where $\hat{\pi}_{01}, \hat{\pi}_{10}$ and $\hat{\pi}_{00}$ are obtained based on the empirical characteristic function and Fourier analysis \cite{jin2007estimating}. If any of $\hat{\pi}_{00}, \hat{\pi}_{10}, \hat{\pi}_{01}$ is close to 1, DACT then follows $U(0,1)$ approximately. Otherwise, the DACT statistic deviates from $U(0,1)$. Under this scenario, the DACT method adapts Efron's empirical null framework \cite{efron2001empirical} to estimate the null distribution of the transformed DACT statistic using inverse standard normal distribution function. The final p-value is calibrated using the empirical null distribution.

However, the reference distribution for the DACT test statistic has not been established. When none of $\pi_{00}, \pi_{10}, \pi_{01}$ is close to 1, although Efron's method has been adapted as a remedy to estimate the null distribution of the transformed DACT statistic, it remains unclear how close the estimation is to the truth. In fact, the cumulative distribution function for the DACT statistic is complicated, because the third term $p_{max,j}^2$ in \eqref{DACT} depends on the larger of the first two terms such that the three terms are dependent. Therefore, DACT should be used cautiously when $\pi_{00},\pi_{01},\pi_{10}$ are all far from 1, for example, when they are all 1/3 say.

\subsubsection{A new variant of HDMT: Sobel-comp}

We propose a variant of HDMT using Sobel's test statistic, called Sobel-comp. Under $H_{01,j}$ and $H_{10,j}$, $T_{sobel,j} \sim N(0,1)$ asymptotically. Under $H_{00,j}$, $T_{sobel,j} \sim N(0,1/4)$ asymptotically. The asymptotic reference distribution for $T_{sobel,j}$ is:
$$(\hat{\pi}_{01}+\hat{\pi}_{10} )N(0,1) + \hat{\pi}_{00} N(0,1/4)$$
\noindent where $\widehat{\pi}_{01}, \widehat{\pi}_{10}, \widehat{\pi}_{00}$ are obtained from the HDMT method. When $|Z_{\beta,j}|>|Z_{\alpha,j}|$, the p-value for HDMT under $H_{00,j}$ is identical no matter how large $|Z_{\beta,j}|$ is. Therefore, the HDMT method loses power since a stronger effect of the mediator on the outcome does not increase the power to detect the mediation effect if the exposure has a relatively weak effect on the mediator. In contrast, the p-value for Sobel-comp under $H_{00,j}$ decreases as $|Z_{\beta,j}|$ increases. In particular,

\textit{Proposition 1.} Suppose $|Z_{\beta,j}| > |Z_{\alpha,j}| \geq 0$. The case-specific p-value under $H_{00,j}$ from Sobel-comp is smaller than that from HDMT if $|Z_{\beta,j}|> max \Big(|Z_{\alpha,j}|, \bigg\{ 4 \bigg(\Phi^{-1} \big(2\Phi(|Z_{\alpha,j}|)^2 \big) \bigg)^{-2} - Z_{\alpha,j}^{-2} \bigg\}^{-1/2}$ \Big), where $\Phi(\cdot)$ is the cumulative distribution function of a standard normal random variable.

\textit{Proposition 1} is also true when we interchange $|Z_{\beta,j}|$ and $|Z_{\alpha,j}|$. The proof of \textit{Proposition 1} is provided in the supplementary materials. However, in addition to the conditions in \textit{Proposition 1}, Sobel-comp requires $\pi_{00}$ close to 1 to be more powerful than HDMT. On the other hand, unlike HDMT which can estimate $P(p_{\alpha,j}<p_{max,j}|H_{10,j})$ and $P(p_{\beta,j}<p_{max,j}|H_{01,j})$ to further increase power with finite samples, it is difficult to extend Sobel-comp using similar technique because $Z_{\alpha,j}$ and $Z_{\beta,j}$ in the Sobel's statistic are not separable.

\subsection{Simulation setup}

We evaluate the performance of Sobel's test, MaxP, JT-comp, HDMT, Sobel-comp and DACT in terms of false positive rate (FPR) under the null hypothesis and true positive rate (TPR) under the alternative hypothesis in simulation scenarios by varying 1) the proportion of the null and the alternative components, denoted as $\pi_{00}, \pi_{01}, \pi_{10}, \pi_{11}$; 2) the sample size $n$; 3) the variation of the non-zero parameters $\alpha$, $\beta$ across mediators; and 4) $R^2$ in the data-generating models. We assess the mediation effect of $J=100,000$ mediators (denoted as $M_j$ where $j \in \{1,2,...,J\}$) from the exposure $(X)$ to the outcome $(Y)$. For the j-th pair of models, we first generate the exposure $X \sim N(0,1)$ and then generate $M_j$ and $Y$ from:
\begin{equation} \label{MX_simu}
    M_j = \alpha_j X + \epsilon_{M_j}
\end{equation}
\begin{equation} \label{YMX_simu}
    Y = \beta_j M_j + \beta_X X + \epsilon_{Y}
\end{equation}
\noindent where $\epsilon_{M_j} \sim N(0,\sigma_{M_j}^2)$, $\epsilon_{Y} \sim N(0,\sigma_{Y}^2)$ and $\beta_X=1$. For $J$ pairs of models, with probability $\pi_{00}$, $\alpha_j = \beta_j =0$; with probability $\pi_{01}$, $\alpha_j=0,\beta_j \sim N(0, \tau^2)$; with probability $\pi_{10}$, $\alpha_j \sim N(0,5\tau^2), \beta_j = 0$; and with probability $\pi_{11}$, $\alpha_j \sim N(0, 5\tau^2), \beta_j \sim N(0,\tau^2)$. The parameter $\tau$ controls the dispersion of the non-zero coefficients. 

To evaluate the FPR for the six methods under the composite null hypothesis, $\pi_{11}$ is set as $0$. We construct four classes of scenarios (\textbf{Table \ref{table: null_simu}}). 
\begin{table} [H]
\centering
\begin{tabular}{ccccc}
    \hline
    & $\pi_{11}, \pi_{01}, \pi_{10},\pi_{00}$ & Sample size & $\tau$ & $R^2$ \\ \hline
Sparse null 1 &  $0, 0.001,0.001,0.998$   &     $(200, 500, 1000)$          &   $(0.1, 0.3, 0.7)$  &           Not controlled                       \\
Dense null 1  &   $0,0.33, 0.33, 0.34$   &      $(200, 500, 1000)$          &   $(0.1, 0.3, 0.7)$  &  Not controlled                               \\
Sparse null 2 &  $0, 0.001,0.001,0.998$    &     $(200, 500, 1000)$          &   $(0.3)$  &           $(0.1,0.15,0.2)$                        \\
Dense null 2  &   $0,0.33, 0.33, 0.34$  &      $(200, 500, 1000)$          &   $(0.3)$  &  $(0.1,0.15,0.2)$                                                           \\
        \hline
    \end{tabular}
        \caption{Simulation scenarios for comparing false positive rates . With probability $\pi_{01}$, $\alpha_j=0$ and $\beta_j \sim N(0,\tau^2)$; with probability $\pi_{10}, \alpha_j \sim N(0,5\tau^2)$ and $\beta_j = 0$; with probability $\pi_{00}, \alpha_j=\beta_j=0$. The last column refers to the $R^2$ in the data generating models.}
    \label{table: null_simu}
\end{table}

In Sparse\&Dense null 1 scenarios, $\delta_{M_j} = \delta_{Y}=1$. In contrast to Sparse\&Dense null 1 scenarios where $R^2$ varies across mediators, Sparse\&Dense null 2 scenarios control $R^2$ at the same level. We calculate the FPR at the nominal significance levels of $10^{-3}, 10^{-4}, 10^{-5}, 10^{-6}$, and $5 \times 10^{-7}$, where $5\times 10^{-7}$ corresponds to controlling the overall FWER at $0.05$. Under the null hypothesis, the FPR given a significance level is calculated as the proportion of p-values among 100,000 tests below this level. We repeat this process 2,000 times and average FPRs over 2,000 replicates.

For power comparison, we follow the same data generation process described above except that we also simulate data under the alternative hypothesis. We have four classes of scenarios in \textbf{Table \ref{table: alternative_simu} }.

\begin{table} [H]
\centering
\begin{tabular}{ccccc}
    \hline
    & $\pi_{11}, \pi_{01}, \pi_{10},\pi_{00}$ & Sample size & $\tau$ & $R^2$ \\ \hline
Sparse alternative 1 &  $0.001, 0.001,0.001,0.997$   &     $(200, 500, 1000)$          &   $(0.1, 0.3, 0.7)$  &           Not controlled                       \\
Dense alternative 1  &   $0.2,0.2, 0.2, 0.4$   &      $(200, 500, 1000)$          &   $(0.1, 0.3, 0.7)$  &  Not controlled                               \\
Sparse alternative 2 &  $0.001, 0.001,0.001,0.997$    &     $(200, 500, 1000)$          &   $(0.3)$  &           $(0.1,0.15,0.2)$                        \\
Dense alternative 2  &   $0.2,0.2, 0.2, 0.4$  &      $(200, 500, 1000)$          &   $(0.3)$  &  $(0.1,0.15,0.2)$                                                           \\
        \hline
    \end{tabular}
        \caption{Simulation scenarios for comparing true positive rates. With probability $\pi_{11}, \alpha_j \sim N(0,5\tau^2), \beta_j \sim N(0,\tau^2)$; with probability $\pi_{01}$, $\alpha_j=0$ and $\beta_j \sim N(0,\tau^2)$; with probability $\pi_{10}, \alpha_j \sim N(0,5\tau^2)$ and $\beta_j = 0$; with probability $\pi_{00}, \alpha_j=\beta_j=0$. The last column refers to the $R^2$ in the data generating models.}
    \label{table: alternative_simu}
\end{table}

Under the control of the true FDR at 0.05, we evaluate the TPR for each method by calculating the number of observed rejections under which the alternative hypothesis is true to the total number true non-null signals. Calculating the true FDR is possible in simulation studies since the underlying truth is known. We repeat the process 200 times, and the TPR is averaged over all 200 replicates. We use existing R software and packages to implement JT-comp (\url{t
http://www.stat.sinica.edu.tw/ythuang/JT-Comp.zip}), DACT \cite{liu2020large} and HDMT \cite{dai2020multiple}. 

\subsection{Data example using MESA: study design and methods}

We apply all six methods (Sobel's test, the MaxP test, JT-comp, HDMT, Sobel-comp and DACT) to study the mediation mechanism of DNA methylation levels at CpG sites in the pathway from adult SES to HbA1c using data from MESA \cite{bild2002multi}. Our exposure, adult SES, defined by educational attainment, is a risk factor for cardiovascular disease and diabetes \cite{whitaker2014association, telfair2012educational}. Our outcome, HbA1c, which reflects the three-month average blood sugar level, is a critical measurement in the diagnosis of diabetes \cite{world2011use} and is a known risk factor for cardiovascular disease \cite{singer1992association,sakurai2013hba1c,yeung2018impact}. 
We assume that the effect direction is from educational attainment to HbA1c level since the exposure has remained unchanged during the study and was collected before measuring HbA1c. Moreover, previous research has reported potential causality between educational attainment and type 2 diabetes \cite{liang2021educational}. In addition, educational attainment is associated with DNAm \cite{van2018dna}, and DNAm is also associated with HbA1c \cite{chen2020dna}. It is thus of interest to identify DNAm sites that mediate the effect of educational attainment on HbA1c. 

Since correlated mediators may lead to inflated Type I error rates and spurious signals, we selected a subset of 228,088 potentially mediating CpGs that were, at most, only weakly correlated with one another. We provide details for processing MESA data in the supplementary materials. For each CpG site, we obtained $z_{\alpha,j}$ and $z_{\beta,j}$ from linear mixed models for testing $\alpha_j=0$ (effect of the exposure on the j-th mediator) and $\beta_j=0$ (effect of the j-th mediator on the outcome). In both models, we adjusted for age, sex and race as potential confounders and adjusted for the estimated proportions of residual non-monocytes (neutrophils, B cells, T cells, and natural killer cells) to account for potential contamination by non-monocyte cell types. We included the methylation chip and position as random effects to account for potential batch effects. In addition, we adjusted for the exposure in the outcome-mediator model. We applied the six mediation methods to the selected 228,088 CpGs, and obtained p-values from each method for testing the mediation effect.  CpG sites with a significant mediation effect are determined by the p-value threshold of $2.19\times10^{-7}$, which corresponds to controlling FWER at 0.05. 

\section{Results}

\subsection{Simulation results}

\subsubsection{False positive rates under the composite null hypothesis}

\begin{figure}[H]
	\centering
	\caption{QQ plots for p-values from Sobel's test, the MaxP test, JT-comp, HDMT, Sobel-comp and DACT under the \textit{Sparse null 1} scenario. $n$ is the sample size. The total number of mediators is 100,000. For $j=1,2,...,100,000$, with probability $\pi_{01}=0.001$, $\alpha_j=0$ and $\beta_j \sim N(0,\tau^2)$; with probability $\pi_{10}=0.001, \alpha_j \sim N(0,5\tau^2)$ and $\beta_j = 0$; with probability $\pi_{00}=0.998, \alpha_j=\beta_j=0$. } 
	\includegraphics[width=1\textwidth]{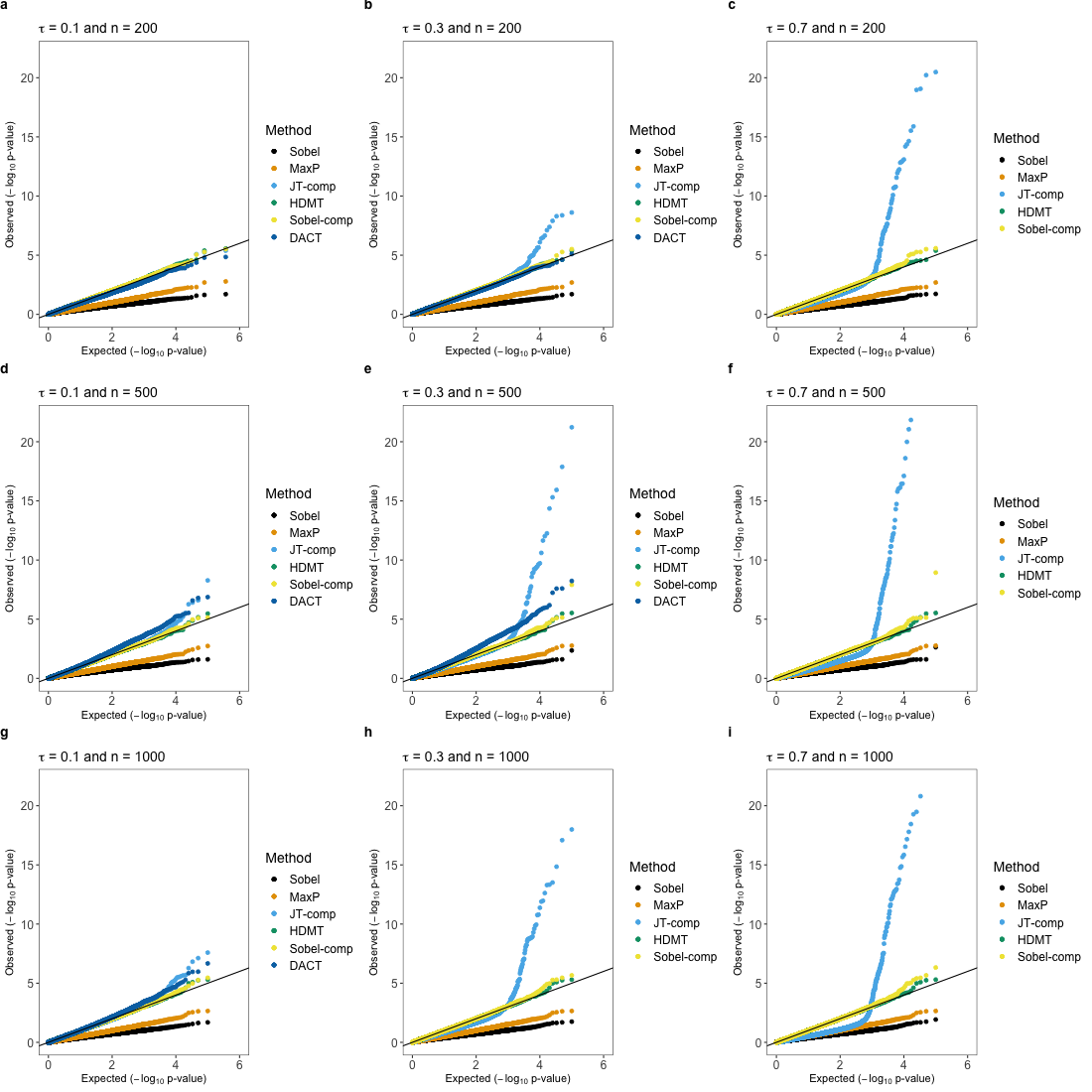}
	\label{qq_typeIerror_sparse}
\end{figure}

In \textbf{Table \ref{table_sparse_null_1}}, we present FPR from six methods under the \textit{Sparse null 1} scenario, where $(\pi_{01}, \pi_{10},\pi_{00}) = (0.001,0.001,0.998)$, the sample size $n \in (200, 500, 1000)$ and $\tau \in (0.1, 0.3, 0.7)$. To better illustrate the distributions of p-values, we provide QQ plots from one replication in \textbf{Figure \ref{qq_typeIerror_sparse}}. 
For all nine cases, Sobel's test is the most conservative test, followed by the MaxP test. P-values from both tests are uniformly larger than the expected p-values due to large $\pi_{00}$. R package DACT fails in certain cases, e.g. when $\tau=0.7$ or when $n=1000$.
When $n=200$ and $\tau=0.1$, the FPRs from HDMT and Sobel-comp are close to expected values at the cut-off higher than $10^{-6}$, but are inflated at a lower cut-off. 
In comparison, p-values from JT-comp and DACT are greatly inflated, especially when the cut-off is lower than $10^{-6}$. At the cut-off of $5\times 10^{-7}$, the ratio of the FPR to the corresponding cut-off for JT-comp, DACT, Sobel-comp, and HDMT is 15.4, 1.5, 2.1 and 21.9, respectively. 
When increasing $n$ from 200 to 1000 with $\tau=0.1$, the FPR for JT-comp dramatically increases. In comparison, Sobel-comp is less inflated and HDMT almost keeps the same level of FPR. Similar trends are observed with an increasing $\tau$.

\begin{figure}[H]
	\centering
	\caption{QQ plots for p-values from Sobel's test, the MaxP test, JT-comp, HDMT, Sobel-comp and DACT under the \textit{Dense null 1} scenario. $n$ is the sample size. The total number of mediators is 100,000. For $j=1,2,...,100,000$, with probability $\pi_{01}=0.33$, $\alpha_j=0$ and $\beta_j \sim N(0,\tau^2)$; with probability $\pi_{10}=0.33, \alpha_j \sim N(0,5\tau^2)$ and $\beta_j=0$; with probability $\pi_{00}=0.34, \alpha_j=\beta_j=0$. } 
	\includegraphics[width=1\textwidth]{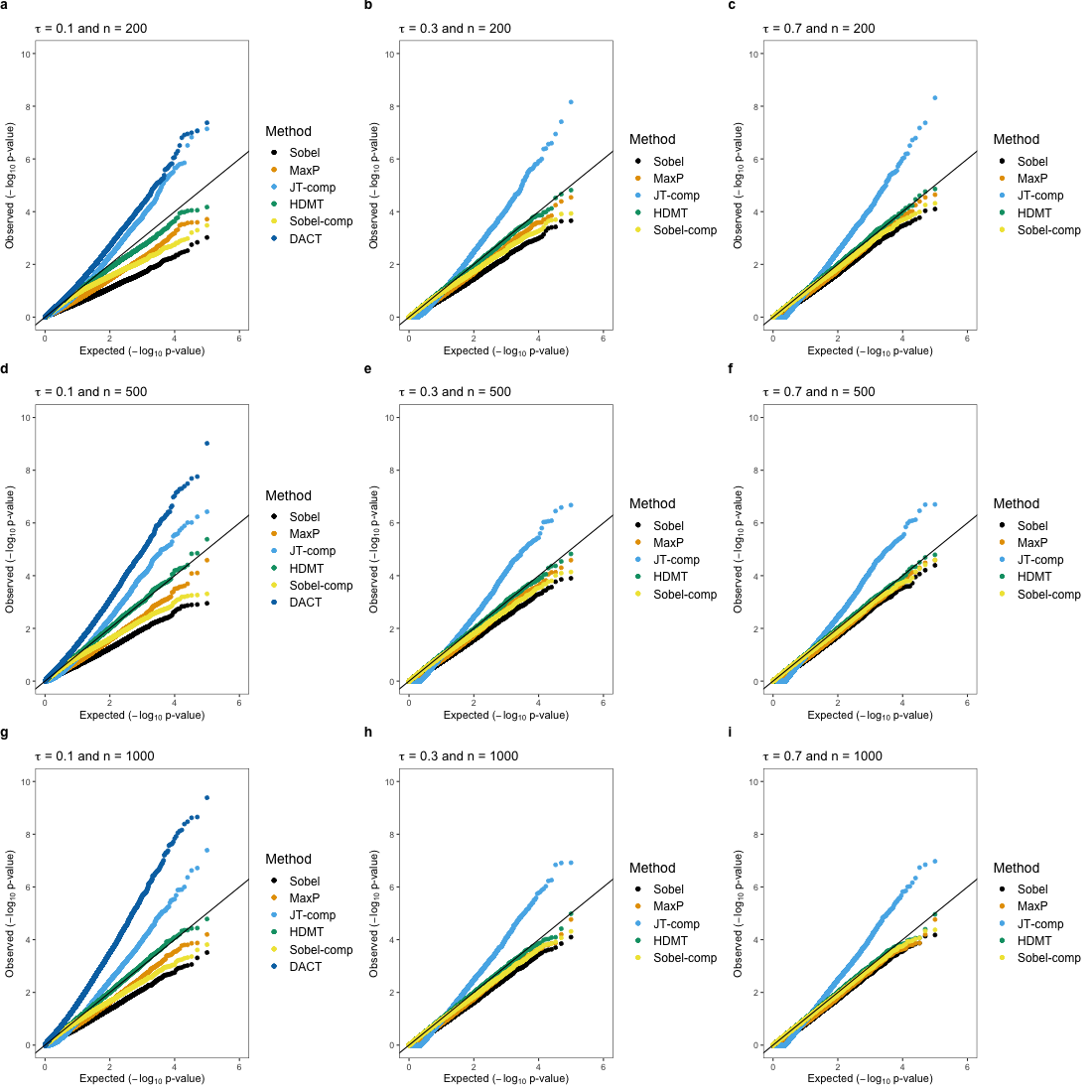}
	\label{qq_typeIerror_dense}
\end{figure}

When the non-zero coefficients are dense in the \textit{Dense null 1} scenario (\textbf{Figure \ref{qq_typeIerror_dense}} and \textbf{Table \ref{table_dense_null_1}}), HDMT is the only method that maintains the FPR at the nominal level in all scenarios, and is robust to the change of $n$ or $\tau$. 

In \textbf{Tables \ref{table_sparse_null_2} and \ref{table_dense_null_2}}, we present the FPR for the \textit{Sparse\&Dense null 2} scenarios, where $R^2 \in (0.1, 0.15, 0.2)$ is controlled across $J$ tests. Overall, the impact of $R^2$ is similar to $\tau$ in the \textit{Sparse\&Dense null 1} scenario for all methods except DACT. In the \textit{Sparse null 2} scenario, increasing $R^2$ ameliorates the inflated FPR for DACT. However, there is no clear trend of how the sample size impacts DACT. With a fixed $R^2$, DACT has the largest FPR when $n=500$, but has smaller ones when $n=200$ and $n=1000$. 

\subsubsection{True positive rates under the alternative hypothesis}

\begin{figure}[H]
	\centering
	\caption{The average true positive rate over 200 replicates when controlling the true false discovery rate (FDR) at 0.05 for Sobel's test, MaxP, JT-comp, HDMT, Sobel-comp and DACT under the \textit{Spase and Dense alternative 1} scenarios. The total number of mediators is 100,000. $n$ is the sample size. For $j=1,2,...,100,000$, with probability $\pi_{11}$, $\alpha_j \sim N(0,5\tau^2), \beta_j \sim N(0,\tau^2)$; with probability $\pi_{01}$, $\alpha_j=0$ and $\beta_j \sim N(0,\tau^2)$; with probability $\pi_{10}, \alpha_j \sim N(0,5\tau^2)$ and $\beta_j = 0$; with probability $\pi_{00}, \alpha_j=\beta_j=0$. Under the \textit{Sparse alternative 1} scenario, $\pi_{11}, \pi_{10}, \pi_{01}, \pi_{00}$ are set as $0.001, 0.001, 0.001, 0.997$ and under the \textit{Dense alternative 1} scenario, $\pi_{11}, \pi_{10}, \pi_{01}, \pi_{00}$ are set as $0.2, 0.2, 0.2, 0.4$.} 
	\includegraphics[width=1\textwidth]{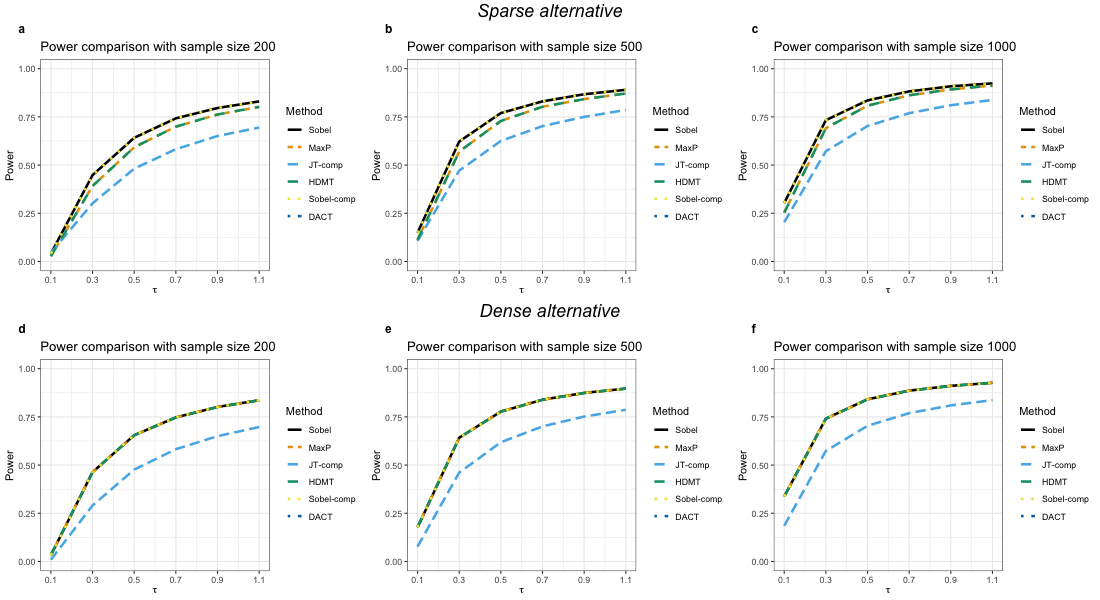}
	\label{power_fdr_1}
\end{figure}

Results of the TPRs using $FDR<0.05$ in \textit{Sparse and Dense alternative 1} scenarios are shown in \textbf{Figure \ref{power_fdr_1}}. DACT fails when $\tau > 0.1$. Under the \textit{Sparse alternative 1} scenario, JT-comp has lower TPR than the four other methods in most simulation scenarios, except when $\tau$ is small (e.g. $\tau = 0.1$) and the sample size is small (e.g. $n=200$). Sobel's test and Sobel-comp have the highest TPRs, closely followed by HDMT and MaxP. The TPR increases for all methods when the sample size increases. Sobel's test and Sobel-comp perform the same because the rank of the weighted composite p-values is unchanged and so are the MaxP test and HDMT. Under the \textit{Dense alternative 1} scenario, the TPR of Sobel's test, MaxP, HDMT and Sobel-comp is the same under the control of FDR. JT-comp has the lowest TPR among all methods.

\begin{figure}[H]
	\centering
	\caption{The average true positive rate over 200 replicates when controlling the true false discovery rate (FDR) at 0.05 for Sobel's test, MaxP, JT-comp, HDMT, Sobel-comp and DACT under the \textit{Spase and Dense alternative 2} scenarios. The total number of mediators is 100,000. $n$ is the sample size. For $j=1,2,...,100,000$, with probability $\pi_{11}$, $\alpha_j \sim N(0,5\times 0.3^2), \beta_j \sim N(0,0.3^2)$; with probability $\pi_{01}$, $\alpha_j=0$ and $\beta_j \sim N(0,0.3^2)$; with probability $\pi_{10}, \alpha_j \sim N(0,5\times 0.3^2)$ and $\beta_j = 0$; with probability $\pi_{00}, \alpha_j=\beta_j=0$. $R^2$ is controlled in the j-th mediator-exposure and outcome-mediator models. Under the \textit{Sparse alternative 2} scenario, $\pi_{11}, \pi_{10}, \pi_{01}, \pi_{00}$ are set as $0.001, 0.001, 0.001, 0.997$ and under the \textit{Dense alternative 2} scenario, $\pi_{11}, \pi_{10}, \pi_{01}, \pi_{00}$ are set as $0.2, 0.2, 0.2, 0.4$.} 
	\includegraphics[width=1\textwidth]{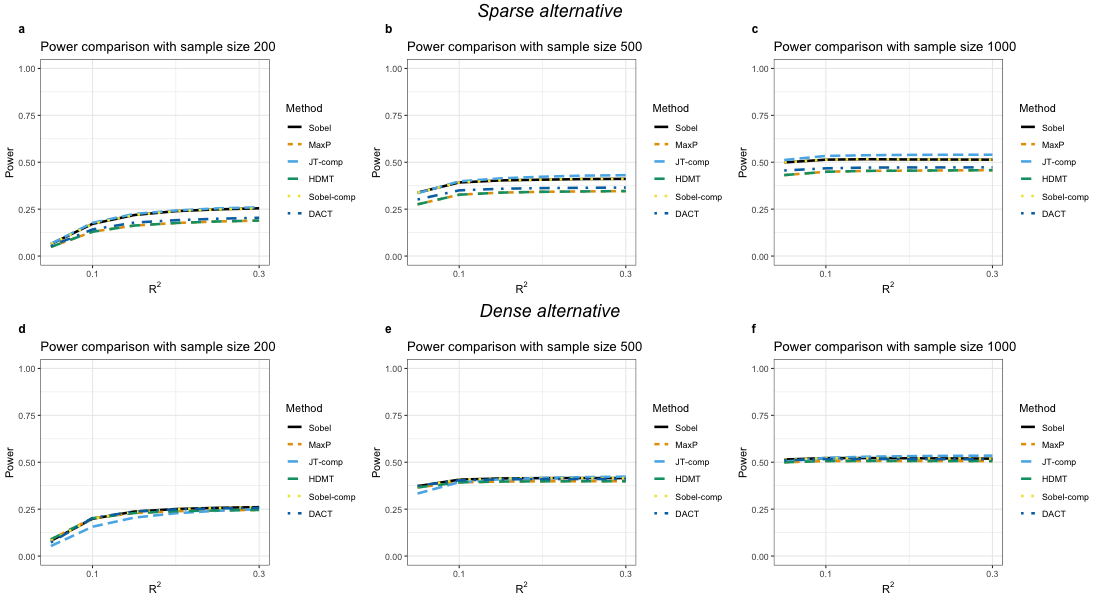}
	\label{power_fdr_2}
\end{figure}

Results for the average TPR using FDR<0.05 in \textit{Sparse and Dense alternative 2} scenarios are shown in \textbf{Figure \ref{power_fdr_2}}. Under the \textit{Sparse alternative 2} scenario, JT-comp, Sobel's test and Sobel-comp have the highest TPRs, followed by the other three methods. The TPR for each methods first increases as $R^2$ increases and stays the same afterward. All methods have increasing TPR as $n$ increases. The TPR for each method is nearly the same with the change of $R^2$ when $n=1000$. Under the \textit{Dense alternative 2} case, all methods have similar TPR and the trend with varying $n$ and $R^2$ is similar to the \textit{Sparse alternative 2} scenario. 

\subsection{Results from MESA}

\begin{figure}[H]
	\centering
	\caption{QQ plot for 6 mediation hypothesis testing methods, including Sobel's test, MaxP, JT-comp, HDMT, Sobel-comp and DACT with 963 observations. The outcome is the continuous HbA1c level, the exposure is the binary adult SES, and the mediators are 228,088 CpG sites. In the mediator-exposure model, we adjusted for age, sex, race and residual white blood cell types (neutrophils, B cells, T cells, and natural killer cells), and included the methylation chip and position as random effects to account for potential batch effects. In addition, we adjusted for the exposure in the outcome-mediator model.} 
	\includegraphics[width=0.8\textwidth]{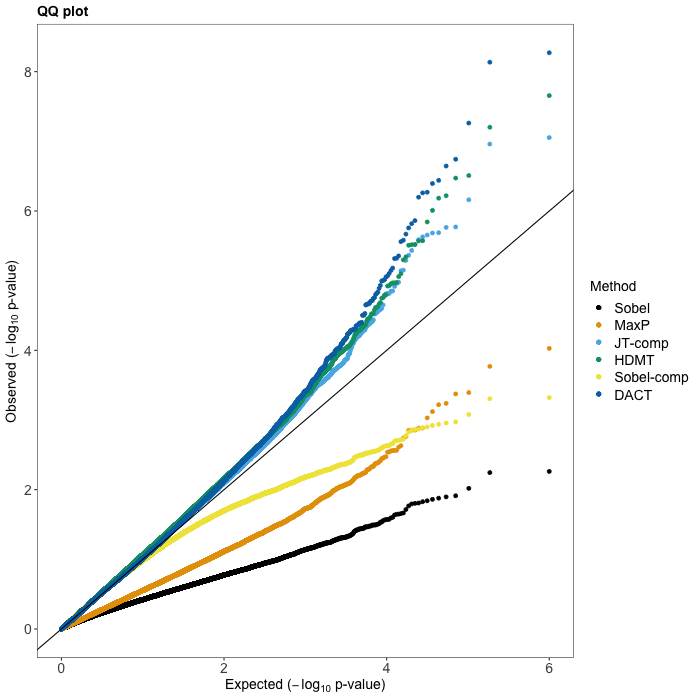}
	\label{qq_plot_mediation}
\end{figure}

In \textbf{Figure \ref{qq_plot_mediation}}, we present the QQ plot for p-values of all 228,088 CpGs from six methods, including Sobel's test, the MaxP test, JT-comp, HDMT, Sobel-comp and DACT. As expected, p-values from Sobel's test and the MaxP test were deflated, potentially due to a large number of zero $\alpha_j$ and $\beta_j$. JT-comp identified two significant CpGs, HDMT identified two significant CpGs, and DACT identified four significant CpGs (\textbf{Table \ref{table_DACT}}). Two CpG sites, cg10508317 and cg01288337, were significant using all three methods (\textbf{Table \ref{table_overlapped}}). In contrast, Sobel-comp detected no significant mediation effects probably because $\hat{\pi}_{00}$ is far from 1 ($\hat{\pi}_{00} = 0.872, \hat{\pi}_{01}=0.032,\hat{\pi}_{10} = 0.052$). 

The CpG site cg10508317 in the SOCS3 gene on chromosome 17 encodes a protein that is involved in the signaling pathways of key hormones such as insulin \cite{pedroso2019socs3}. It has been found that increased SOCS3 expression is associated with insulin resistance \cite{pedroso2019socs3}, which is directly related to HbA1c. The CpG site cg01288337 is in the RIN3 gene on chromosome 14. The RIN3 gene encodes a member of the RIN family of Ras interaction-interference proteins and is next to the SLC24A4 gene. Recent studies showed that SLC24A4/RIN3 is significantly associated with brain glucose metabolism in humans \cite{stage2016effect} and SLC24A4 knockout mice revealed brain glucose hypometabolism \cite{li2014essential}.

\begin{table}[ht]
\scriptsize
\captionsetup[table]{labelsep=space, justification=raggedright, singlelinecheck=off}
\begin{threeparttable}

\caption{: Two mediation pathways identified by JT-comp, HDMT and DACT after controlling the FWER at $0.05$. The exposure is adult SES and the outcome is HbA1c. The total number of mediators is 228,088. In the mediator-exposure model, we adjusted for age, sex, race and residual white blood cell types (neutrophils, B cells, T cells, and natural killer cells), and included the methylation chip and position as random effects to account for potential batch effects. In addition, we adjusted for the exposure in the outcome-mediator model. The estimated mediation effect is $\hat{\alpha}\hat{\beta}$ and the proportion of mediation effect is provided in the parenthesis. The 95\% confidence interval (CI) is calculated based on 1,000 bootstrap samples.}
\begin{tabular}{ccccccccccc}
  \hline
CpG & Chr & Gene & \begin{tabular}[c]{@{}c@{}}UCSC RefGene\\ Group\end{tabular} & $\hat{\alpha}$ & $\hat{\beta}$& \begin{tabular}[c]{@{}c@{}}Mediation effect\\ (proportion)\end{tabular} & 95\% CI & $p_{JT-comp}$ & $p_{HDMT}$ & $p_{DACT}$\\ 
  \hline
cg10508317 & 17 & SOCS3 & Body  & -9.92E-02 & -1.78E-01 & 0.018 (0.18) & (0.009,0.029)	 & 8.84E-08 & 6.28E-08 & 5.35E-09 \\ 
  cg01288337 & 14 & RIN3 & Body & 5.55E-02 & 3.06E-01 & 0.017 (0.17)  & (0.009,0.028) & 1.09E-07 & 2.21E-08 & 7.32E-09 \\ 
   \hline
\end{tabular}
\label{table_overlapped}
\end{threeparttable}

\end{table}

\newpage

\section{Discussion}

We reviewed and compared the testing performance of six mediation methods (Sobel's test, MaxP, JT-comp, HDMT, DACT and Sobel-comp). Our study indicates that the methods which use the mixture reference distribution (HDMT, Sobel-comp) can better control false positive rates and yield larger true positive rates. However, there is no uniform dominance of one method over the others across all simulation scenarios. The performance of the methods differs according to values of $\pi_{00},\pi_{01}, \pi_{10},\pi_{11}$, the sample size and the strength of independent variables explaining the variation of the dependent variable in the two models \eqref{YMX_model} and \eqref{MX_model}, as captured by the variance of non-zero $\alpha,\beta$ or $R^2$ in models \eqref{MX_simu} and \eqref{YMX_simu}.

Under the null hypothesis, the distribution of p-values is strongly affected by the three proportions, $\pi_{00},\pi_{01}, \pi_{10}$, for all methods except HDMT. Our simulation studies show that HDMT is the only method that controls FPR when non-zero coefficients are dense, i.e. when $\pi_{01}$ and $\pi_{10}$ are large. On the other hand, when non-zero coefficients are sparse, Sobel-comp performs similar to HDMT. In comparison, JT-comp maintains the nominal level of FPR only when the sample size is small and the variances of non-zero $\alpha$ and $\beta$ are small (or $R^2$ is small). The application of JT-comp is limited to sparse settings with small samples and relatively weak signals. 

\begin{figure}[H]
	\centering
	\caption{Decision tree for choosing the optimal mediation hypothesis testing method based on the simulation studies for the normally-distributed outcome. $\pi_{11}, \pi_{01}, \pi_{10},\pi_{00}$ are the proportion of $(\alpha\neq 0,\beta \neq 0), (\alpha = 0,\beta \neq 0),(\alpha \neq 0,\beta = 0)$, and $(\alpha=\beta=0)$, respectively. } 
	\includegraphics[width=1\textwidth]{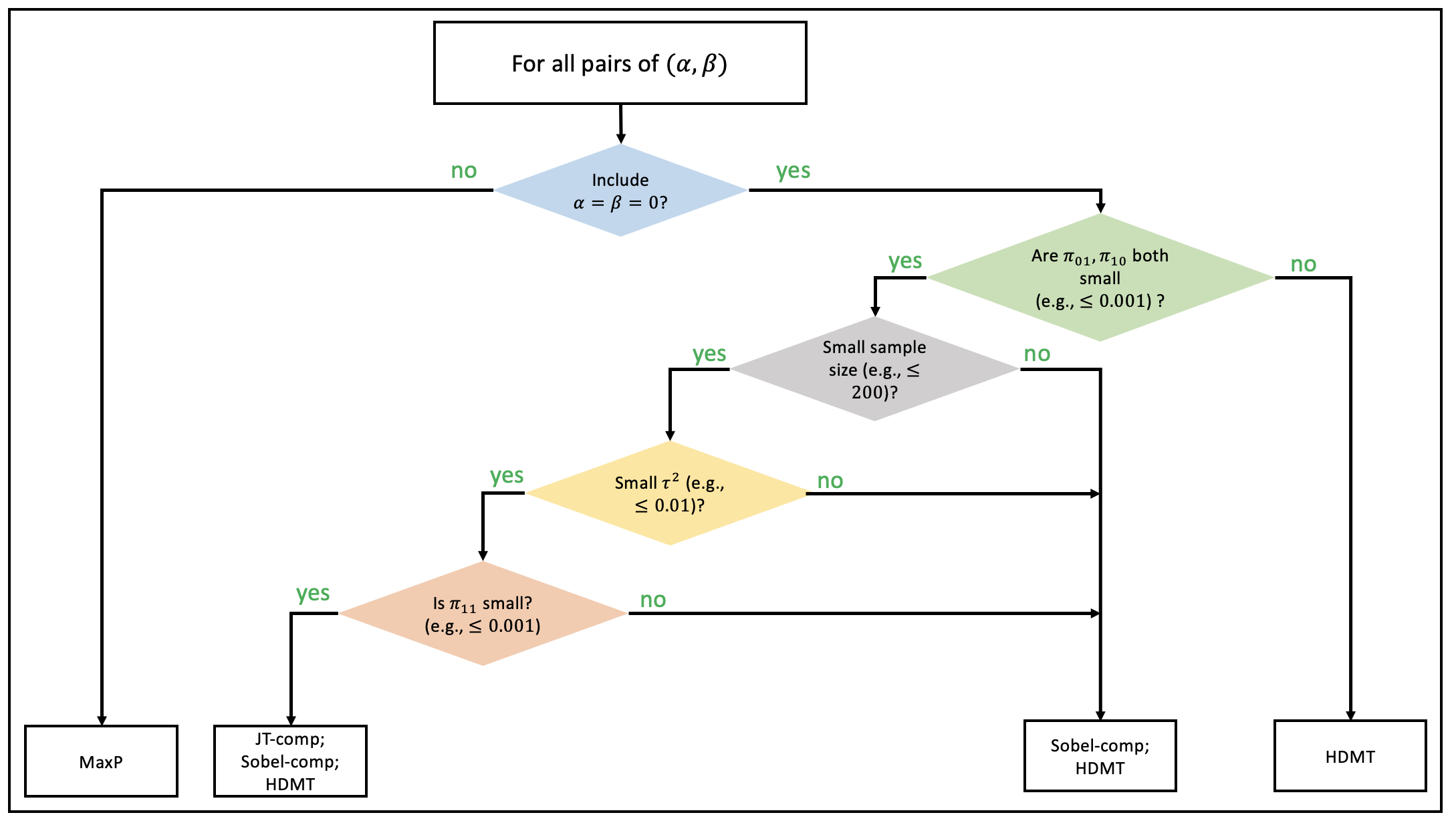}
	\label{decision_tree}
\end{figure}

Under the alternative hypothesis with sparse signals, all methods perform similar with a small sample size $n$ and small $\tau$. As $n$ and $\tau$ increase, Sobel-comp is most powerful method with the greatest TPR, followed by HDMT. Under the dense settings, Sobel-comp has the same TPR as HDMT. In practice, we recommend to first estimate $\pi_{01},\pi_{10},\pi_{00}$ using R package \texttt{HDMT} \cite{dai2020multiple} and then choose the method based on $\hat{\pi}_{01},\hat{\pi}_{10},\hat{\pi}_{00}$. Sobel-comp is preferred when $\hat{\pi}_{01}$ and $\hat{\pi}_{10}$ are close to 0. Otherwise, HDMT is preferred. Although we do not provide strict guidelines, our simulation studies show that when $\pi_{01} = \pi_{10} = \pi_{11} = 0.001$, Sobel-comp is the most powerful method in almost all scenarios. We summarize key features, advantages and limitations for all the six methods in \textbf{Table \ref{table:summaryOfMethods}} and provide a decision tree for choosing an appropriate method in \textbf{Figure \ref{decision_tree}}. 

\begin{sidewaystable}
 \caption{A summary of methods with key advantages and limitations. For $J$ tests, $\pi_{11}, \pi_{01}, \pi_{10},\pi_{00}$ are the true proportion of $(\alpha\neq 0,\beta \neq 0), (\alpha = 0,\beta \neq 0),(\alpha \neq 0,\beta = 0)$, and $(\alpha=\beta=0)$, respectively. $\tau^2$ is the variance for the non-zero $\alpha$, $\beta$. $p_\alpha$ and $p_\beta$ are the two-sided p-value for $Z_\alpha$ and $Z_\beta$, respectively.}
\label{table:summaryOfMethods}
\footnotesize
    \begin{tabularx}{\columnwidth}{smmbbBB} \Xhline{2\arrayrulewidth} \hline \hline
   \textbf{Method}     &  \begin{tabular}[c]{@{}c@{}} \textbf{First Author} \\\textbf{Year} \\ \textbf{(Reference No.)}\end{tabular}  &  \begin{tabular}[c]{@{}c@{}} \textbf{R package} \\\textbf{ (package name)} \end{tabular} & \textbf{Test statistic} & \begin{tabular}[c]{@{}c@{}} \textbf{Reference} \\\textbf{Distribution} \end{tabular}  & \textbf{Advantages}     & \textbf{Limitations}  \\ \Xhline{2\arrayrulewidth} \hline
   
   Sobel's test & Sobel, 1982 \cite{sobel1982asymptotic} & \xmark & $\frac{Z_\alpha}{\sqrt{1+(Z_\alpha/Z_\beta)^2}}$ & $N(0,1)$ & Protect the false positive rate (FPR) at the nominal level under the null hypothesis. Robust for any value of $\pi_{00},\pi_{01},\pi_{10}$ and $\tau$.                             & Conservative since the multivariate delta method fails when $\alpha$ and $\beta$ are zero.\\ \hline
   
   MaxP  & \begin{tabular}[c]{@{}l@{}}Mackinnon, \\ 2002 \cite{mackinnon2002comparison}\end{tabular} & \xmark & $max(p_\alpha,p_\beta)$ & $U(0,1)$ & Protect the FPR at the nominal level under the null hypothesis. Robust for any value of $\pi_{00},\pi_{01},\pi_{10}$ and $\tau$. Uniformly more powerful than Sobel's test. & Conservative because of using the incorrect reference distribution when $\alpha$ and $\beta$ are zero.  \\ \Xhline{2\arrayrulewidth} \hline
   
   JT-comp  & Huang, 2019 \cite{huang2019genome} & \cmark & $Z_\alpha Z_\beta$ &
    $H_{00}$: Standard normal product distribution.
    $H_{01}, H_{10}$: Normal product distribution with non-zero mean.

   & Correct mixture reference distribution for $Z_\alpha Z_\beta$. No need to estimate $\pi_{00},\pi_{01},\pi_{10}$. Keeping the FPR close to the nominal level when the sample size and $\tau$ are small. & Inflated FPR at a small significance threshold. Inflated FPR when the sample size or $\tau$ increases. Only works with small samples and relatively weak signals. \\ \hline
   
   HDMT     &  Dai, 2020 \cite{dai2020multiple}  & \cmark (\texttt{HDMT}) & $max(p_\alpha,p_\beta)$ & $H_{00}: Beta(2,1)$. $H_{01}, H_{10}: U(0,1)$.  & Correct mixture reference distribution for the MaxP test statistic. Maintaining the FPR close to the nominal level for any value of $\pi_{00},\pi_{01},\pi_{10}$ and $\tau$. More powerful than Sobel's test and the MaxP test. Provides finite-sample size adjustment for p-values to increase power. & Power lose when $p_{\beta}$ is much smaller than $p_{\alpha}$ and vice versa.\\ \hline 
   
   Sobel-comp & - & \xmark & $\frac{Z_\alpha}{\sqrt{1+(Z_\alpha/Z_\beta)^2}}$ & $H_{00}: N(0,1/4)$. $H_{01},H_{10}: N(0,1)$ & Correct mixture reference distribution for Sobel's test statistic. Maintaining the FPR close to the nominal level and more powerful than HDMT when $\pi_{01}$ and $\pi_{10}$ are close to 0.
   & Conservative if $\pi_{01}$ or $\pi_{10}$ is far from 0 due to the use of the asymptotic reference distribution under $H_{01}$ and $H_{10}$.\\ \Xhline{2\arrayrulewidth} \hline
      
   DACT   &  Liu, 2020 \cite{liu2020large}  & \cmark (\texttt{DACT}) & $\widehat{\pi}_{01} p_\alpha + \widehat{\pi}_{10} p_\beta + \widehat{\pi}_{00} p_{max}^2$ & $U(0,1)$ approximately. & Weights the case-specific p-values to construct a composite test statistic to accommodate the composite nature of the null hypothesis. & Exact reference distribution of $DACT$ statistic is not established. Approximation of the reference distribution is often inaccurate, causing the FPR deviating from the nominal level.\\ \hline  \hline \Xhline{2\arrayrulewidth} 
   
    \end{tabularx}
\end{sidewaystable}

A common limitation for all six methods is that none of them work when mediators are correlated. Presented with correlated mediators, univariate mediation analysis does not adjust for all the mediator-outcome confounders affected by the exposure, resulting in a violation of assumption 4 mentioned in Section 1. In this case, it is necessary to extend the mediation analysis models to jointly account for multiple correlated mediators \cite{song2020bayesian, song2020bayesian2}. For computational reasons, we only explore a range of parameters. Parameter values beyond this range combined with correlated mediators are of interest for future analysis.

The two significant CpGs we identified in the SOCS3 and RIN3 genes from MESA add to a growing body of literature for the mediating role of DNA methylation between socioeconomic status and disease risk factors associated with HbA1c \cite{song2020bayesian,giurgescu2019neighborhood}. However, a limitation of our analysis is that our mediator (methylation) and outcome (HbA1c) were measured concurrently. Therefore, we identify statistical mediation but are unable to formally evaluate causal mediation. More studies are needed to fully understand the underlying biological mechanisms that link socioeconomic disadvantage to HbA1c-associated diseases. 

(word count: 3984)

\section*{Acknowledgments}

MESA and the MESA SHARe project are conducted and supported by the National Heart, Lung, and Blood Institute (NHLBI) in collaboration with MESA investigators. Support for MESA is provided by contracts 75N92020D00001, HHSN268201500003I, N01-HC-95159, 75N92020D00005, N01-HC-95160, 75N92020D00002, N01-HC-95161, 75N92020D00003, N01-HC-95162, 75N92020D00006, N01-HC-95163, 75N92020D00004, N01-HC-95164, 75N92020D00007, N01-HC-95165, N01-HC-95166, N01-HC-95167, N01-HC-95168, N01-HC-95169, UL1-TR-000040, UL1-TR-001079, UL1-TR-001420, UL1-TR-001881, and DK063491. The MESA Epigenomics \& Transcriptomics Studies were funded by NIH grants 1R01HL101250, 1RF1AG054474, R01HL126477, R01DK101921, and R01HL135009. The analysis for this study was funded by NHLBI (R01HL141292). 

\section*{Conflict of Interest}

All authors declared no conflict of interest.

\newpage

\printbibliography

\newpage

\section*{Supplementary Materials}
\beginsupplement

\noindent \textbf{Proof of proposition 1}

\begin{proof}
In this proof, we drop the subscript $j$ since the statement is true for all $j \in \{1,2,...,J \}$. When $|Z_\beta|>|Z_\alpha|$, under $H_{00}$,
\begin{equation*}
    T_{sobel} = \frac{|Z_\alpha|}{\sqrt{1+Z_\alpha^2/Z_\beta^2}} \sim N(0,1/4)
\end{equation*}
\begin{equation*}
    p_{max} = 2\Phi(|Z_\alpha|) \sim Beta(2,1)
\end{equation*}
The p-value for Sobel-comp under $H_{00}$ is:
\begin{align*}
    p_{sobel}^{00} &= 2 \int_{\frac{|Z\alpha|}{\sqrt{1+Z_\alpha^2/Z_\beta^2}}}^{\infty} (2\pi\frac{1}{4})^{-1/2} exp\big(-2u^2 \big) du \\
    &=2\int_{\frac{2|Z_\alpha|}{\sqrt{1+Z_\alpha^2/Z_\beta^2}}}^{\infty} (2\pi)^{-1/2}exp \big(-1/2v^2 \big)dv \\
    &= 2\Phi \bigg(\frac{2|Z_\alpha|}{\sqrt{1+Z_\alpha^2/Z_\beta^2}} \bigg)
\end{align*}
The p-value for HDMT under $H_{00}$ is:
\begin{equation*}
    p_{max}^2 = 4\Phi(|Z_\alpha|)^2
\end{equation*}
Notice that $p_{sobel}^{00}$ is a decreasing function of $|Z_\beta|$, and 
\begin{align*}
    & p_{sobel}^{00}<p_{max}^2 \\
    \Longleftrightarrow\ \ \   & 2\Phi \bigg(\frac{2|Z_\alpha|}{\sqrt{1+Z_\alpha^2/Z_\beta^2}} \bigg) < 4\Phi(|Z_\alpha|)^2\\
    \Longleftrightarrow\ \ \ & \frac{2|Z_\alpha|}{\sqrt{1+Z_\alpha^2/Z_\beta^2}} >\Phi^{-1} \bigg(2\Phi(|Z_\beta|)^2 \bigg)\\
    \Longleftrightarrow\ \ \ & |Z_\beta|> \bigg\{ 4 \bigg(\Phi^{-1} \big(2\Phi(|Z_\alpha|)^2 \big) \bigg)^{-2} - Z_\alpha^{-2} \bigg\}^{-1/2}
\end{align*}
Note that $\bigg\{ 4 \bigg(\Phi^{-1} \big(2\Phi(|Z_\alpha|)^2 \big) \bigg)^{-2} - Z_\alpha^{-2} \bigg\}>0$. Therefore, a sufficient condition for $p_{sobel}^{00}<p_{max}^2$ is $|Z_\beta|> max \Big(|Z_\alpha|, \bigg\{ 4 \bigg(\Phi^{-1} \big(2\Phi(|Z_\alpha|)^2 \big) \bigg)^{-2} - Z_\alpha^{-2} \bigg\}^{-1/2} \Big)$.

\end{proof}

\newpage

\noindent \textbf{Detailed description of MESA data}

MESA is a population-based longitudinal study designed to investigate the predictors and progression of subclinical cardiovascular disease in a cohort of 6,814 participants \cite{bild2002multi}. Clinical, socio-demographic, lifestyle and behavior, laboratory, nutrition and medication data have been collected at multiple examinations beginning in 2000-2002. We used participants' educational attainment based on their highest degree at MESA Exam 1 as a measure of adult SES (less than a 4-year college degree as 1 vs. with a 4-year college degree or higher as 0). DNAm levels were measured using the Illumina Infinium HumanMethylation450 Beadchip on purified monocytes from a random subsample of 1,264 non-Hispanic white, African-American, and Hispanic MESA participants between April 2010 and February 2012 (corresponding to MESA Exam 5). A total of 402,339 CpGs remained after quality control and filtering, including: ``detected'' methylation levels in <90\% of MESA samples using a detection p-value cut-off of 0.05,  overlap with a repetitive element or region, presence of SNPs within 10 base pairs according to Illumina annotation, non-reliable probes recommended by DMRcate (having SNPs with minor allele frequency $>0.05$ within 2 base pairs or cross reactive probes), probes on sex chromosomes, SNPs, and other non-CpG targeting probes. Additional details about the data collection and processing procedures can be found in \cite{liu2013methylomics}. We used HbA1c measured at Exam 5 as the outcome. Our analysis focused on the participants taking no insulin or oral hypoglycemic medication. After removing missing values, a total of 963 individuals remained for analysis.

For the j-th CpG site, where $j=1,2,...,402339$, we obtained $z_{\alpha,j}$ and $z_{\beta,j}$ from linear mixed models for testing $\alpha_j=0$ (effect of the exposure on the j-th mediator) and $\beta_j=0$ (effect of the j-th mediator on the outcome). In both models, we adjusted for age, sex and race as potential confounders and adjusted for the estimated proportions of residual non-monocytes (neutrophils, B cells, T cells, and natural killer cells) to account for potential contamination by non-monocyte cell types. We included the methylation chip and position as random effects to account for potential batch effects. In addition, we adjusted for the exposure in the outcome-mediator model. 

Before performing mediation analysis, since correlated mediators may lead to inflated Type I error rates and spurious signals, we selected a subset of 228,088 potentially mediating CpGs that were, at most, only weakly correlated with one another (correlation coefficient $\leq 0.3$). More specifically, we first calculated the correlation matrix for all CpGs on each chromosome. Then we found the mediator with the smallest MaxP p-value. Next, we identified and removed the group of mediators which were correlated with this mediator with correlation coefficient larger than 0.3. We repeated the previous steps until all elements in the correlation matrix were less than or equal to 0.3.

\newpage

\begin{table}[ht]
\centering
{\scriptsize
\caption{The mean and standard deviation of the ratio of the false positive rates to the nominal significance level based on 2,000 replications using Sobel's test, MaxP, JT-comp, HDMT, Sobel-comp and DACT under the \textit{Sparse null 1} case. $n$ is the sample size. The total number of mediators is 100,000. With probability $\pi_{01}=0.001$, $\alpha=0$ and $\beta \sim N(0,\tau^2)$; with probability $\pi_{10}=0.001, \alpha \sim N(0,5\tau^2)$ and $\beta = 0$; with probability $\pi_{00}=0.998, \alpha=\beta=0$. }
\label{table_sparse_null_1}
\begin{tabular}{lllllll}
\hline
Cut-off & Sobel & MaxP & JT-comp & HDMT & Sobel-comp & DACT \\ 
  \hline
  \multicolumn{7}{c}{$n=200, \tau = 0.1$}    \\ \hline
  
$10^{-3}$ & 0.00 (0.00) & 0.00 (0.00) & 1.11 (0.10) & 1.06 (0.11) & 0.90 (0.34) & 1.46 (1.78) \\ 
  $10^{-4}$ & 0.00 (0.00) & 0.00 (0.01) & 1.49 (0.38) & 1.14 (0.36) & 0.83 (0.67) & 2.85 (4.10) \\ 
  $10^{-5}$ & 0.00 (0.00) & 0.00 (0.00) & 3.06 (1.67) & 1.26 (1.12) & 1.03 (1.30) & 6.36 (10.29) \\ 
  $10^{-6}$ &  0.00 (0.00) &  0.00 (0.00) & 10.02 (10.09) &  1.47 (3.94) &  1.75 (4.53) & 16.09 (30.78) \\ 
  $5 \times 10^{-7}$ &  0.00 (0.00) &  0.00 (0.00) & 15.40 (17.71) &  1.53 (5.68) &  2.18 (6.85) & 21.94 (45.68) \\ 
   \hline
   
   \multicolumn{7}{c}{$n=200, \tau = 0.3$}    \\ \hline
   $10^{-3}$ & 0.00 (0.00) & 0.00 (0.01) & 1.26 (0.11) & 1.07 (0.12) & 0.91 (0.36) & 1.60 (1.82) \\ 
  $10^{-4}$ & 0.00 (0.00) & 0.00 (0.01) & 3.40 (0.57) & 1.18 (0.38) & 0.95 (0.80) & 3.21 (4.29) \\ 
  $10^{-5}$ &  0.00 (0.00) &  0.00 (0.02) & 17.27 (4.13) &  1.39 (1.21) &  1.67 (1.96) &  7.29 (11.02) \\ 
  $10^{-6}$ &   0.00 (0.00) &   0.00 (0.00) & 108.76 (33.03) &   2.05 (4.62) &   4.87 (8.36) &  18.62 (31.79) \\ 
  $5 \times 10^{-7}$ &   0.00 (0.00) &   0.00 (0.00) & 191.40 (62.55) &   2.35 (7.04) &   7.34 (14.11) &  25.11 (45.02) \\ 
   \hline
   
   \multicolumn{7}{c}{$n=200, \tau = 0.7$}    \\ \hline
   $10^{-3}$ & 0.00 (0.00) & 0.00 (0.01) & 1.10 (0.09) & 1.02 (0.16) & 0.93 (0.37) &   NA (NA) \\ 
  $10^{-4}$ & 0.00 (0.01) & 0.00 (0.02) & 6.79 (0.73) & 1.03 (0.47) & 1.10 (0.90) &   NA (NA) \\ 
  $10^{-5}$ &  0.00 (0.02) &  0.00 (0.05) & 52.94 (6.71) &  1.15 (1.26) &  2.44 (2.76) &    NA (NA) \\ 
  $10^{-6}$ &   0.00 (0.00) &   0.01 (0.32) & 426.53 (60.62) &   1.76 (4.53) &   9.17 (13.32) &     NA (NA) \\ 
  $5 \times 10^{-7}$ &   0.00 (0.00) &   0.01 (0.45) & 801.93 (115.88) &   2.20 (7.07) &  14.30 (22.35) &     NA (NA) \\ 
   \hline
   
   \multicolumn{7}{c}{$n=500, \tau = 0.1$}    \\ \hline
   $10^{-3}$ & 0.00 (0.00) & 0.00 (0.00) & 1.14 (0.10) & 1.05 (0.11) & 0.97 (0.29) & 1.13 (1.62) \\ 
  $10^{-4}$ & 0.00 (0.00) & 0.00 (0.01) & 2.08 (0.45) & 1.11 (0.34) & 1.04 (0.62) & 2.27 (3.75) \\ 
  $10^{-5}$ & 0.00 (0.00) & 0.00 (0.03) & 7.50 (2.70) & 1.26 (1.12) & 1.63 (1.58) & 5.74 (10.89) \\ 
  $10^{-6}$ &  0.00 (0.00) &  0.00 (0.00) & 38.94 (19.27) &  1.65 (4.09) &  4.13 (6.85) & 20.82 (61.49) \\ 
  $5 \times 10^{-7}$ &  0.00 (0.00) &  0.00 (0.00) & 65.99 (35.02) &  1.96 (6.15) &  5.92 (11.29) & 33.65 (114.42) \\ 
   \hline
   
   \multicolumn{7}{c}{$n=500, \tau = 0.3$}    \\ \hline
   $10^{-3}$ & 0.00 (0.00) & 0.00 (0.01) & 1.18 (0.10) & 1.03 (0.14) & 1.01 (0.30) &   NA (NA) \\ 
  $10^{-4}$ & 0.00 (0.01) & 0.00 (0.01) & 5.43 (0.68) & 1.08 (0.41) & 1.29 (0.76) &   NA (NA) \\ 
  $10^{-5}$ &  0.00 (0.00) &  0.00 (0.04) & 38.43 (5.85) &  1.28 (1.23) &  3.01 (2.48) &    NA (NA) \\ 
  $10^{-6}$ &   0.00 (0.00) &   0.00 (0.00) & 292.64 (50.35) &   2.05 (4.64) &  11.13 (12.48) &     NA (NA) \\ 
  $5 \times 10^{-7}$ &   0.00 (0.00) &   0.00 (0.00) & 542.37 (97.17) &   2.48 (7.01) &  17.43 (21.55) &     NA (NA) \\ 
   \hline
   
   \multicolumn{7}{c}{$n=500, \tau = 0.7$}    \\ \hline
   $10^{-3}$ & 0.00 (0.00) & 0.00 (0.01) & 1.03 (0.09) & 1.03 (0.15) & 1.03 (0.31) &   NA (NA) \\ 
  $10^{-4}$ & 0.00 (0.01) & 0.00 (0.01) & 8.10 (0.82) & 1.07 (0.43) & 1.45 (0.83) &   NA (NA) \\ 
  $10^{-5}$ &  0.00 (0.03) &  0.00 (0.04) & 67.27 (7.26) &  1.30 (1.26) &  3.88 (3.04) &    NA (NA) \\ 
  $10^{-6}$ &   0.00 (0.00) &   0.00 (0.00) & 564.57 (66.14) &   2.16 (4.78) &  16.38 (16.36) &     NA (NA) \\ 
  $5 \times 10^{-7}$ &    0.00 (0.00) &    0.00 (0.00) & 1072.33 (128.19) &    2.71 (7.44) &   25.90 (27.85) &      NA (NA) \\ 
   \hline
   
   \multicolumn{7}{c}{$n=1000, \tau = 0.1$}    \\ \hline
   $10^{-3}$ & 0.00 (0.00) & 0.00 (0.00) & 1.19 (0.11) & 1.04 (0.11) & 1.00 (0.26) &   NA (NA) \\ 
  $10^{-4}$ & 0.00 (0.00) & 0.00 (0.01) & 2.94 (0.54) & 1.11 (0.33) & 1.15 (0.61) &   NA (NA) \\ 
  $10^{-5}$ &  0.00 (0.00) &  0.00 (0.03) & 14.52 (3.82) &  1.35 (1.17) &  2.12 (1.79) &    NA (NA) \\ 
  $10^{-6}$ &  0.00 (0.00) &  0.00 (0.00) & 90.51 (29.89) &  1.91 (4.45) &  6.72 (9.08) &    NA (NA) \\ 
  $5 \times 10^{-7}$ &   0.00 (0.00) &   0.00 (0.00) & 160.19 (56.83) &   2.32 (6.92) &   9.97 (15.33) &     NA (NA) \\ 
   \hline
   
   \multicolumn{7}{c}{$n=1000, \tau = 0.3$}    \\ \hline
   $10^{-3}$ & 0.00 (0.00) & 0.00 (0.01) & 1.09 (0.09) & 1.03 (0.14) & 1.05 (0.28) &   NA (NA) \\ 
  $10^{-4}$ & 0.00 (0.01) & 0.00 (0.01) & 6.66 (0.73) & 1.07 (0.40) & 1.43 (0.73) &   NA (NA) \\ 
  $10^{-5}$ &  0.00 (0.02) &  0.00 (0.04) & 51.77 (6.56) &  1.35 (1.26) &  3.63 (2.66) &    NA (NA) \\ 
  $10^{-6}$ &   0.00 (0.00) &   0.00 (0.22) & 417.38 (58.22) &   2.11 (4.70) &  14.93 (14.85) &     NA (NA) \\ 
  $5 \times 10^{-7}$ &   0.00 (0.00) &   0.01 (0.45) & 783.81 (113.24) &   2.65 (7.48) &  23.61 (25.70) &     NA (NA) \\ 
   \hline
   
   \multicolumn{7}{c}{$n=1000, \tau = 0.7$}    \\ \hline
   $10^{-3}$ & 0.00 (0.00) & 0.00 (0.01) & 1.06 (0.09) & 1.03 (0.14) & 1.07 (0.28) &   NA (NA) \\ 
  $10^{-4}$ & 0.00 (0.01) & 0.00 (0.01) & 8.84 (0.84) & 1.08 (0.41) & 1.56 (0.79) &   NA (NA) \\ 
  $10^{-5}$ &  0.00 (0.04) &  0.00 (0.04) & 74.83 (7.54) &  1.39 (1.28) &  4.39 (3.07) &    NA (NA) \\ 
  $10^{-6}$ &   0.00 (0.00) &   0.00 (0.22) & 638.24 (68.79) &   2.28 (4.89) &  19.36 (17.50) &     NA (NA) \\ 
  $5 \times 10^{-7}$ &    0.00 (0.00) &    0.01 (0.45) & 1218.12 (133.97) &    2.90 (7.77) &   31.27 (30.74) &      NA (NA) \\ 
   \hline
\end{tabular}
}
\end{table}

\newpage
\begin{table}[ht]
\centering
{\scriptsize
\caption{The mean and standard deviation of the ratio of the false positive rates to the nominal significance level based on 2,000 replications using Sobel's test, MaxP, JT-comp, HDMT, Sobel-comp and DACT under the \textit{Dense null 1} case. $n$ is the sample size. The total number of mediators is 100,000. With probability $\pi_{01}=0.33$, $\alpha=0$ and $\beta \sim N(0,\tau^2)$; with probability $\pi_{10}=0.33, \alpha \sim N(0,5\tau^2)$ and $\beta = 0$; with probability $\pi_{00}=0.34, \alpha=\beta=0$. }
\label{table_dense_null_1}
\begin{tabular}{lllllll}
  \hline
Cut-off & Sobel & MaxP & JT-comp & HDMT & Sobel-comp & DACT \\ 
  \hline
  \multicolumn{7}{c}{$n=200, \tau = 0.1$}    \\ \hline
$10^{-3}$ & 0.01 (0.01) & 0.14 (0.04) & 3.48 (0.18) & 1.08 (0.10) & 0.04 (0.02) & 7.16 (2.17) \\ 
  $10^{-4}$ &  0.00 (0.01) &  0.11 (0.11) &  7.92 (0.89) &  1.16 (0.34) &  0.01 (0.03) & 18.32 (7.32) \\ 
  $10^{-5}$ &  0.00 (0.02) &  0.12 (0.34) & 18.47 (4.25) &  1.29 (1.16) &  0.00 (0.06) & 49.82 (25.09) \\ 
  $10^{-6}$ &   0.00 (0.00) &   0.08 (0.86) &  42.73 (20.33) &   1.66 (4.07) &   0.00 (0.00) & 142.94 (88.76) \\ 
  $5 \times 10^{-7}$ &   0.00 (0.00) &   0.09 (1.34) &  54.96 (32.89) &   1.86 (6.01) &   0.00 (0.00) & 196.51 (130.30) \\ 
   \hline
   
   \multicolumn{7}{c}{$n=200, \tau = 0.3$}    \\ \hline

   $10^{-3}$ & 0.13 (0.04) & 0.46 (0.07) & 4.60 (0.21) & 0.86 (0.10) & 0.27 (0.05) &   NA (NA) \\ 
  $10^{-4}$ &  0.09 (0.09) &  0.47 (0.22) & 11.52 (1.08) &  0.86 (0.30) &  0.18 (0.13) &    NA (NA) \\ 
  $10^{-5}$ &  0.07 (0.25) &  0.55 (0.74) & 29.43 (5.53) &  0.94 (0.97) &  0.14 (0.38) &    NA (NA) \\ 
  $10^{-6}$ &  0.04 (0.63) &  0.53 (2.35) & 76.89 (27.70) &  1.02 (3.23) &  0.09 (0.94) &    NA (NA) \\ 
  $5 \times 10^{-7}$ &   0.05 (1.00) &   0.52 (3.31) & 101.87 (44.87) &   1.01 (4.60) &   0.08 (1.26) &     NA (NA) \\ 
   \hline

   \multicolumn{7}{c}{$n=200, \tau = 0.7$}    \\ \hline
   $10^{-3}$ & 0.35 (0.06) & 0.63 (0.08) & 4.75 (0.21) & 1.03 (0.10) & 0.59 (0.08) &   NA (NA) \\ 
  $10^{-4}$ &  0.30 (0.17) &  0.69 (0.26) & 12.03 (1.10) &  1.12 (0.33) &  0.51 (0.22) &    NA (NA) \\ 
  $10^{-5}$ &  0.30 (0.55) &  0.87 (0.92) & 31.19 (5.67) &  1.33 (1.15) &  0.50 (0.70) &    NA (NA) \\ 
  $10^{-6}$ &  0.26 (1.58) &  0.91 (3.11) & 82.34 (28.33) &  1.53 (4.02) &  0.41 (1.97) &    NA (NA) \\ 
  $5 \times 10^{-7}$ &   0.28 (2.35) &   0.88 (4.29) & 109.64 (46.28) &   1.54 (5.77) &   0.40 (2.80) &     NA (NA) \\ 
   \hline
   
   \multicolumn{7}{c}{$n=500, \tau = 0.1$}    \\ \hline

   $10^{-3}$ &  0.03 (0.02) &  0.24 (0.05) &  4.16 (0.20) &  1.04 (0.10) &  0.09 (0.03) & 13.78 (2.37) \\ 
  $10^{-4}$ &  0.01 (0.03) &  0.21 (0.14) &  9.99 (0.98) &  1.08 (0.34) &  0.04 (0.06) & 41.14 (9.28) \\ 
  $10^{-5}$ &   0.00 (0.06) &   0.17 (0.41) &  24.32 (4.87) &   1.12 (1.08) &   0.01 (0.12) & 127.45 (36.18) \\ 
  $10^{-6}$ &   0.00 (0.00) &   0.14 (1.18) &  60.01 (24.74) &   1.08 (3.25) &   0.00 (0.00) & 406.34 (144.92) \\ 
  $5 \times 10^{-7}$ &   0.00 (0.00) &   0.10 (1.41) &  79.58 (40.30) &   1.18 (4.88) &   0.00 (0.00) & 578.58 (222.45) \\ 
   \hline

   \multicolumn{7}{c}{$n=500, \tau = 0.3$}    \\ \hline

   $10^{-3}$ & 0.22 (0.05) & 0.51 (0.07) & 4.69 (0.21) & 0.88 (0.09) & 0.40 (0.06) &   NA (NA) \\ 
  $10^{-4}$ &  0.15 (0.13) &  0.51 (0.22) & 11.78 (1.06) &  0.87 (0.30) &  0.28 (0.17) &    NA (NA) \\ 
  $10^{-5}$ &  0.11 (0.33) &  0.48 (0.70) & 30.04 (5.44) &  0.84 (0.91) &  0.19 (0.44) &    NA (NA) \\ 
  $10^{-6}$ &  0.06 (0.80) &  0.56 (2.40) & 77.33 (27.88) &  0.88 (3.03) &  0.12 (1.11) &    NA (NA) \\ 
  $5 \times 10^{-7}$ &   0.09 (1.34) &   0.54 (3.30) & 103.02 (45.89) &   0.91 (4.36) &   0.11 (1.48) &     NA (NA) \\ 
   \hline

   \multicolumn{7}{c}{$n=500, \tau = 0.7$}    \\ \hline
   
   $10^{-3}$ & 0.42 (0.06) & 0.62 (0.08) & 4.75 (0.21) & 0.98 (0.10) & 0.68 (0.08) &   NA (NA) \\ 
  $10^{-4}$ &  0.36 (0.19) &  0.64 (0.25) & 11.98 (1.06) &  1.01 (0.32) &  0.60 (0.25) &    NA (NA) \\ 
  $10^{-5}$ &  0.31 (0.56) &  0.63 (0.80) & 30.81 (5.53) &  1.03 (1.01) &  0.51 (0.71) &    NA (NA) \\ 
  $10^{-6}$ &  0.30 (1.74) &  0.70 (2.64) & 79.39 (28.47) &  1.04 (3.29) &  0.47 (2.19) &    NA (NA) \\ 
  $5 \times 10^{-7}$ &   0.21 (2.04) &   0.71 (3.76) & 106.34 (46.60) &   1.11 (4.71) &   0.42 (2.87) &     NA (NA) \\ 
   \hline

   \multicolumn{7}{c}{$n=1000, \tau = 0.1$}    \\ \hline
   
   $10^{-3}$ & 0.07 (0.03) & 0.33 (0.06) & 4.44 (0.20) & 1.03 (0.10) & 0.17 (0.04) &   NA (NA) \\ 
  $10^{-4}$ &  0.03 (0.06) &  0.29 (0.17) & 10.91 (1.03) &  1.05 (0.32) &  0.08 (0.09) &    NA (NA) \\ 
  $10^{-5}$ &  0.02 (0.13) &  0.26 (0.50) & 27.16 (5.29) &  1.09 (1.03) &  0.04 (0.21) &    NA (NA) \\ 
  $10^{-6}$ &  0.02 (0.39) &  0.24 (1.55) & 68.32 (26.56) &  1.12 (3.39) &  0.03 (0.50) &    NA (NA) \\ 
  $5 \times 10^{-7}$ &  0.02 (0.63) &  0.22 (2.09) & 90.70 (43.41) &  1.17 (4.90) &  0.03 (0.77) &    NA (NA) \\ 
   \hline

   \multicolumn{7}{c}{$n=1000, \tau = 0.3$}    \\ \hline
   
   $10^{-3}$ & 0.29 (0.05) & 0.55 (0.07) & 4.73 (0.21) & 0.91 (0.09) & 0.51 (0.07) &   NA (NA) \\ 
  $10^{-4}$ &  0.23 (0.15) &  0.55 (0.23) & 11.89 (1.09) &  0.91 (0.31) &  0.40 (0.20) &    NA (NA) \\ 
  $10^{-5}$ &  0.18 (0.41) &  0.55 (0.74) & 30.29 (5.61) &  0.91 (0.95) &  0.30 (0.55) &    NA (NA) \\ 
  $10^{-6}$ &  0.11 (1.04) &  0.53 (2.29) & 77.89 (28.14) &  0.95 (3.06) &  0.19 (1.37) &    NA (NA) \\ 
  $5 \times 10^{-7}$ &   0.14 (1.67) &   0.52 (3.25) & 103.73 (46.64) &   0.84 (4.06) &   0.18 (1.89) &     NA (NA) \\ 
   \hline

   \multicolumn{7}{c}{$n=1000, \tau = 0.7$}    \\ \hline
   
   $10^{-3}$ & 0.47 (0.07) & 0.63 (0.08) & 4.76 (0.21) & 0.98 (0.10) & 0.76 (0.09) &   NA (NA) \\ 
  $10^{-4}$ &  0.43 (0.20) &  0.64 (0.25) & 12.00 (1.10) &  1.00 (0.32) &  0.69 (0.26) &    NA (NA) \\ 
  $10^{-5}$ &  0.39 (0.61) &  0.65 (0.81) & 30.60 (5.65) &  1.03 (1.02) &  0.63 (0.79) &    NA (NA) \\ 
  $10^{-6}$ &  0.32 (1.76) &  0.64 (2.53) & 79.30 (28.23) &  1.06 (3.27) &  0.55 (2.32) &    NA (NA) \\ 
  $5 \times 10^{-7}$ &   0.31 (2.47) &   0.69 (3.71) & 104.99 (46.75) &   0.99 (4.48) &   0.46 (3.00) &     NA (NA) \\ 
   \hline
\end{tabular}
}
\end{table}

\newpage
\begin{table}[ht]
\centering
{\scriptsize
\caption{The mean and standard deviation of the ratio of the false positive rates to the nominal significance level based on 2,000 replications using Sobel's test, MaxP, JT-comp, HDMT, Sobel-comp and DACT under the \textit{Sparse null 2} case. $n$ is the sample size. The total number of mediators is 100,000. With probability $\pi_{01}=0.001$, $\alpha=0$ and $\beta \sim N(0,0.3^2)$; with probability $\pi_{10}=0.001, \alpha \sim N(0,5\times 0.3^2)$ and $\beta = 0$; with probability $\pi_{00}=0.998, \alpha=\beta=0$. $R^2$ is controlled in the mediator-exposure and outcome-mediator models.}
\label{table_sparse_null_2}
\begin{tabular}{lllllll}
  \hline
Cut-off & Sobel & MaxP & JT-comp & HDMT & Sobel-comp & DACT \\ 
  \hline
  \multicolumn{7}{c}{$n=200, R^2 = 0.1$}    \\ \hline
$10^{-3}$ & 0.00 (0.00) & 0.00 (0.01) & 1.20 (0.11) & 1.06 (0.12) & 0.90 (0.35) & 1.58 (1.83) \\ 
  $10^{-4}$ & 0.00 (0.00) & 0.00 (0.01) & 2.02 (0.46) & 1.16 (0.38) & 0.89 (0.76) & 3.21 (4.37) \\ 
  $10^{-5}$ & 0.00 (0.00) & 0.00 (0.03) & 5.11 (2.36) & 1.36 (1.21) & 1.37 (1.73) & 7.76 (12.62) \\ 
  $10^{-6}$ &  0.00 (0.00) &  0.00 (0.00) & 16.45 (13.27) &  1.85 (4.48) &  2.99 (6.13) & 22.41 (54.88) \\ 
  $5 \times 10^{-7}$ &  0.00 (0.00) &  0.00 (0.00) & 23.71 (22.35) &  2.17 (6.83) &  4.17 (9.82) & 31.85 (91.44) \\ 
   \hline
   
   \multicolumn{7}{c}{$n=200, R^2 = 0.15$}    \\ \hline

$10^{-3}$ & 0.00 (0.00) & 0.00 (0.01) & 1.27 (0.11) & 1.06 (0.12) & 0.91 (0.36) & 1.59 (1.83) \\ 
  $10^{-4}$ & 0.00 (0.00) & 0.00 (0.01) & 2.76 (0.55) & 1.17 (0.38) & 0.94 (0.79) & 3.22 (4.38) \\ 
  $10^{-5}$ & 0.00 (0.00) & 0.00 (0.04) & 9.97 (3.44) & 1.39 (1.22) & 1.65 (2.00) & 7.80 (13.07) \\ 
  $10^{-6}$ &  0.00 (0.00) &  0.00 (0.22) & 42.82 (22.12) &  1.98 (4.60) &  4.44 (7.88) & 23.25 (70.65) \\ 
  $5 \times 10^{-7}$ &  0.00 (0.00) &  0.00 (0.00) & 67.14 (38.99) &  2.34 (7.02) &  6.38 (12.50) & 34.41 (129.76) \\ 
   \hline

   \multicolumn{7}{c}{$n=200, R^2 = 0.2$}    \\ \hline
$10^{-3}$ & 0.00 (0.00) & 0.00 (0.01) & 1.32 (0.11) & 1.06 (0.12) & 0.91 (0.36) & 1.61 (1.84) \\ 
  $10^{-4}$ & 0.00 (0.00) & 0.00 (0.01) & 3.46 (0.60) & 1.17 (0.38) & 0.97 (0.81) & 3.24 (4.38) \\ 
  $10^{-5}$ &  0.00 (0.00) &  0.00 (0.04) & 15.43 (4.35) &  1.40 (1.22) &  1.84 (2.18) &  7.57 (11.98) \\ 
  $10^{-6}$ &  0.00 (0.00) &  0.00 (0.22) & 79.90 (31.26) &  2.00 (4.62) &  5.64 (9.37) & 20.57 (52.19) \\ 
  $5 \times 10^{-7}$ &   0.00 (0.00) &   0.00 (0.00) & 130.40 (55.40) &   2.38 (7.10) &   8.26 (15.14) &  29.00 (93.88) \\ 
   \hline
   
   \multicolumn{7}{c}{$n=500, R^2 = 0.1$}    \\ \hline

$10^{-3}$ & 0.00 (0.00) & 0.00 (0.00) & 1.30 (0.11) & 1.05 (0.11) & 1.00 (0.29) & 1.27 (1.63) \\ 
  $10^{-4}$ & 0.00 (0.00) & 0.00 (0.01) & 3.61 (0.59) & 1.14 (0.35) & 1.20 (0.69) & 2.85 (4.24) \\ 
  $10^{-5}$ &  0.00 (0.00) &  0.00 (0.03) & 17.37 (4.26) &  1.35 (1.16) &  2.50 (2.12) & 10.53 (23.75) \\ 
  $10^{-6}$ &  0.00 (0.00) &  0.00 (0.00) & 93.96 (31.59) &  1.93 (4.28) &  8.26 (10.03) & 66.33 (223.61) \\ 
  $5 \times 10^{-7}$ &   0.00 (0.00) &   0.00 (0.00) & 156.65 (57.45) &   2.40 (6.80) &  12.51 (17.02) & 124.35 (446.18) \\ 
   \hline

   \multicolumn{7}{c}{$n=500, R^2 = 0.15$}    \\ \hline

$10^{-3}$ & 0.00 (0.00) & 0.00 (0.00) & 1.34 (0.11) & 1.05 (0.11) & 1.00 (0.30) & 1.23 (1.66) \\ 
  $10^{-4}$ & 0.00 (0.00) & 0.00 (0.01) & 4.63 (0.66) & 1.14 (0.35) & 1.24 (0.72) & 2.57 (4.04) \\ 
  $10^{-5}$ &  0.00 (0.00) &  0.00 (0.03) & 27.08 (5.19) &  1.37 (1.18) &  2.78 (2.29) &  7.53 (17.31) \\ 
  $10^{-6}$ &   0.00 (0.00) &   0.00 (0.00) & 173.38 (41.99) &   2.07 (4.48) &   9.87 (11.31) &  35.27 (148.51) \\ 
  $5 \times 10^{-7}$ &   0.00 (0.00) &   0.00 (0.00) & 301.01 (79.72) &   2.57 (7.07) &  15.22 (19.09) &  62.57 (295.70) \\ 
   \hline

   \multicolumn{7}{c}{$n=500, R^2 = 0.2$}    \\ \hline
   
$10^{-3}$ & 0.00 (0.00) & 0.00 (0.00) & 1.34 (0.10) & 1.06 (0.11) & 1.01 (0.30) & 1.23 (1.67) \\ 
  $10^{-4}$ & 0.00 (0.00) & 0.00 (0.01) & 5.37 (0.69) & 1.15 (0.35) & 1.28 (0.74) & 2.47 (3.96) \\ 
  $10^{-5}$ &  0.00 (0.00) &  0.00 (0.03) & 34.99 (5.75) &  1.39 (1.19) &  2.94 (2.40) &  6.14 (12.83) \\ 
  $10^{-6}$ &   0.00 (0.00) &   0.00 (0.00) & 244.80 (48.87) &   2.11 (4.56) &  11.00 (12.19) &  20.89 (88.39) \\ 
  $5 \times 10^{-7}$ &   0.00 (0.00) &   0.00 (0.00) & 439.05 (94.04) &   2.65 (7.18) &  17.01 (20.63) &  33.05 (173.35) \\ 
   \hline

   \multicolumn{7}{c}{$n=1000, R^2 = 0.1$}    \\ \hline
   
$10^{-3}$ & 0.00 (0.00) & 0.00 (0.00) & 1.34 (0.10) & 1.05 (0.11) & 1.03 (0.27) & 1.14 (1.59) \\ 
  $10^{-4}$ & 0.00 (0.00) & 0.00 (0.01) & 5.13 (0.69) & 1.13 (0.34) & 1.32 (0.68) & 2.36 (3.86) \\ 
  $10^{-5}$ &  0.00 (0.02) &  0.00 (0.04) & 32.49 (5.64) &  1.40 (1.20) &  3.01 (2.30) &  6.91 (16.64) \\ 
  $10^{-6}$ &   0.00 (0.00) &   0.00 (0.22) & 221.03 (46.60) &   2.05 (4.61) &  11.15 (12.16) &  33.20 (145.14) \\ 
  $5 \times 10^{-7}$ &   0.00 (0.00) &   0.01 (0.45) & 392.65 (88.03) &   2.55 (7.27) &  17.08 (20.52) &  59.07 (289.05) \\ 
   \hline

   \multicolumn{7}{c}{$n=1000, R^2 = 0.15$}    \\ \hline
   
$10^{-3}$ & 0.00 (0.00) & 0.00 (0.00) & 1.32 (0.10) & 1.05 (0.11) & 1.04 (0.27) & 1.13 (1.62) \\ 
  $10^{-4}$ & 0.00 (0.01) & 0.00 (0.01) & 6.07 (0.74) & 1.14 (0.34) & 1.36 (0.70) & 2.22 (3.74) \\ 
  $10^{-5}$ &  0.00 (0.02) &  0.00 (0.04) & 43.05 (6.25) &  1.43 (1.21) &  3.30 (2.45) &  5.10 (10.19) \\ 
  $10^{-6}$ &   0.00 (0.00) &   0.00 (0.22) & 321.66 (55.73) &   2.15 (4.71) &  12.68 (13.18) &  14.34 (51.36) \\ 
  $5 \times 10^{-7}$ &   0.00 (0.00) &   0.01 (0.45) & 588.52 (105.67) &   2.69 (7.47) &  19.84 (22.59) &  20.69 (95.92) \\ 
   \hline

   \multicolumn{7}{c}{$n=1000, R^2 = 0.2$}    \\ \hline
   
$10^{-3}$ & 0.00 (0.00) & 0.00 (0.01) & 1.28 (0.09) & 1.05 (0.11) & 1.04 (0.27) & 1.15 (1.63) \\ 
  $10^{-4}$ & 0.00 (0.01) & 0.00 (0.01) & 6.73 (0.76) & 1.14 (0.34) & 1.39 (0.71) & 2.26 (3.75) \\ 
  $10^{-5}$ &  0.00 (0.02) &  0.00 (0.04) & 50.70 (6.65) &  1.45 (1.22) &  3.48 (2.57) &  5.01 (9.17) \\ 
  $10^{-6}$ &   0.00 (0.00) &   0.00 (0.22) & 397.30 (59.78) &   2.27 (4.83) &  13.80 (14.02) &  12.86 (26.81) \\ 
  $5 \times 10^{-7}$ &   0.00 (0.00) &   0.01 (0.45) & 737.90 (116.72) &   2.82 (7.62) &  21.51 (23.94) &  17.11 (37.74) \\ 
   \hline
\end{tabular}
}
\end{table}

\newpage
\begin{table}[ht]
\centering
{\scriptsize
\caption{The mean and standard deviation of the ratio of the false positive rates to the nominal significance level based on 2000 replications using Sobel's test, MaxP test, JT-comp, HDMT, Sobel-comp and DACT under the \textit{Sparse null 2} case. $n$ is the sample size. The total number of mediators is 100,000. With probability $\pi_{01}=0.33$, $\alpha=0$ and $\beta \sim N(0,0.3^2)$; with probability $\pi_{10}=0.33, \alpha \sim N(0,5\times 0.3^2)$ and $\beta = 0$; with probability $\pi_{00}=0.34, \alpha=\beta=0$. $R^2$ is controlled in the mediator-exposure and outcome-mediator models.}
\label{table_dense_null_2}
\begin{tabular}{lllllll}
  \hline
Cut-off & Sobel & MaxP & JT-comp & HDMT & Sobel-comp & DACT \\ 
  \hline
  \multicolumn{7}{c}{$n=200, R^2 = 0.1$}    \\ \hline
$10^{-3}$ & 0.01 (0.01) & 0.37 (0.06) & 0.78 (0.09) & 1.12 (0.11) & 0.06 (0.03) & 1.06 (0.50) \\ 
  $10^{-4}$ & 0.00 (0.01) & 0.35 (0.19) & 0.33 (0.18) & 1.26 (0.36) & 0.01 (0.03) & 1.32 (0.94) \\ 
  $10^{-5}$ & 0.00 (0.00) & 0.37 (0.61) & 0.12 (0.35) & 1.51 (1.24) & 0.00 (0.02) & 1.73 (1.98) \\ 
  $10^{-6}$ & 0.00 (0.00) & 0.35 (1.87) & 0.05 (0.67) & 1.92 (4.37) & 0.00 (0.00) & 2.39 (5.39) \\ 
  $5 \times 10^{-7}$ & 0.00 (0.00) & 0.27 (2.31) & 0.02 (0.63) & 2.05 (6.39) & 0.00 (0.00) & 2.79 (8.10) \\ 
   \hline
   
   \multicolumn{7}{c}{$n=200, R^2 = 0.15$}    \\ \hline

$10^{-3}$ & 0.04 (0.02) & 0.42 (0.07) & 0.69 (0.09) & 1.13 (0.11) & 0.14 (0.05) & 3.18 (1.06) \\ 
  $10^{-4}$ & 0.01 (0.03) & 0.45 (0.21) & 0.23 (0.15) & 1.27 (0.35) & 0.03 (0.06) & 6.05 (2.76) \\ 
  $10^{-5}$ &  0.00 (0.04) &  0.54 (0.75) &  0.06 (0.24) &  1.51 (1.22) &  0.01 (0.08) & 12.24 (7.61) \\ 
  $10^{-6}$ &  0.00 (0.00) &  0.66 (2.53) &  0.01 (0.32) &  1.96 (4.40) &  0.00 (0.00) & 25.58 (23.17) \\ 
  $5 \times 10^{-7}$ &  0.00 (0.00) &  0.64 (3.58) &  0.01 (0.45) &  2.24 (6.74) &  0.00 (0.00) & 31.95 (33.88) \\ 
   \hline

   \multicolumn{7}{c}{$n=200, R^2 = 0.2$}    \\ \hline
$10^{-3}$ & 0.09 (0.03) & 0.44 (0.07) & 0.63 (0.08) & 1.13 (0.11) & 0.23 (0.06) & 5.58 (1.47) \\ 
  $10^{-4}$ &  0.03 (0.05) &  0.47 (0.21) &  0.17 (0.13) &  1.27 (0.35) &  0.08 (0.09) & 12.54 (4.42) \\ 
  $10^{-5}$ &  0.01 (0.08) &  0.60 (0.79) &  0.04 (0.19) &  1.52 (1.22) &  0.03 (0.16) & 29.89 (13.90) \\ 
  $10^{-6}$ &  0.00 (0.00) &  0.76 (2.72) &  0.00 (0.22) &  1.96 (4.39) &  0.00 (0.00) & 74.23 (45.75) \\ 
  $5 \times 10^{-7}$ &  0.00 (0.00) &  0.78 (3.92) &  0.00 (0.00) &  2.21 (6.73) &  0.00 (0.00) & 99.05 (67.99) \\ 
   \hline
   
   \multicolumn{7}{c}{$n=500, R^2 = 0.1$}    \\ \hline

$10^{-3}$ & 0.09 (0.03) & 0.40 (0.06) & 0.54 (0.07) & 1.05 (0.10) & 0.23 (0.05) & 6.12 (1.56) \\ 
  $10^{-4}$ &  0.02 (0.05) &  0.40 (0.20) &  0.11 (0.11) &  1.09 (0.33) &  0.07 (0.09) & 13.78 (4.71) \\ 
  $10^{-5}$ &  0.01 (0.09) &  0.41 (0.65) &  0.02 (0.14) &  1.17 (1.07) &  0.02 (0.13) & 32.60 (14.54) \\ 
  $10^{-6}$ &  0.00 (0.22) &  0.38 (1.94) &  0.00 (0.22) &  1.15 (3.43) &  0.01 (0.32) & 78.78 (47.59) \\ 
  $5 \times 10^{-7}$ &   0.00 (0.00) &   0.43 (2.90) &   0.00 (0.00) &   1.16 (4.89) &   0.01 (0.45) & 103.77 (69.89) \\ 
   \hline

   \multicolumn{7}{c}{$n=500, R^2 = 0.15$}    \\ \hline

$10^{-3}$ &  0.16 (0.04) &  0.43 (0.07) &  0.49 (0.07) &  1.05 (0.10) &  0.34 (0.06) & 10.31 (1.98) \\ 
  $10^{-4}$ &  0.07 (0.09) &  0.43 (0.20) &  0.08 (0.09) &  1.09 (0.33) &  0.18 (0.13) & 26.86 (6.84) \\ 
  $10^{-5}$ &  0.03 (0.17) &  0.43 (0.66) &  0.01 (0.10) &  1.16 (1.07) &  0.07 (0.27) & 72.74 (23.79) \\ 
  $10^{-6}$ &   0.02 (0.45) &   0.40 (1.97) &   0.00 (0.00) &   1.15 (3.43) &   0.03 (0.55) & 202.01 (85.94) \\ 
  $5 \times 10^{-7}$ &   0.01 (0.45) &   0.45 (2.97) &   0.00 (0.00) &   1.14 (4.81) &   0.04 (0.89) & 275.85 (127.94) \\ 
   \hline

   \multicolumn{7}{c}{$n=500, R^2 = 0.2$}    \\ \hline
   
$10^{-3}$ &  0.21 (0.05) &  0.46 (0.07) &  0.46 (0.07) &  1.05 (0.10) &  0.41 (0.07) & 13.65 (2.14) \\ 
  $10^{-4}$ &  0.13 (0.11) &  0.45 (0.21) &  0.06 (0.08) &  1.09 (0.33) &  0.26 (0.16) & 38.47 (7.96) \\ 
  $10^{-5}$ &   0.06 (0.26) &   0.45 (0.67) &   0.01 (0.09) &   1.17 (1.06) &   0.13 (0.36) & 111.80 (29.46) \\ 
  $10^{-6}$ &   0.04 (0.59) &   0.43 (2.07) &   0.00 (0.00) &   1.15 (3.42) &   0.06 (0.77) & 333.32 (112.87) \\ 
  $5 \times 10^{-7}$ &   0.04 (0.89) &   0.48 (3.13) &   0.00 (0.00) &   1.17 (4.90) &   0.07 (1.18) & 464.10 (171.83) \\ 
   \hline

   \multicolumn{7}{c}{$n=1000, R^2 = 0.1$}    \\ \hline
   
$10^{-3}$ &  0.18 (0.04) &  0.45 (0.07) &  0.45 (0.07) &  1.03 (0.10) &  0.38 (0.06) & 12.49 (2.16) \\ 
  $10^{-4}$ &  0.11 (0.10) &  0.43 (0.21) &  0.06 (0.08) &  1.06 (0.33) &  0.22 (0.15) & 34.29 (7.83) \\ 
  $10^{-5}$ &  0.05 (0.22) &  0.42 (0.65) &  0.00 (0.07) &  1.12 (1.06) &  0.11 (0.34) & 97.28 (27.95) \\ 
  $10^{-6}$ &   0.01 (0.32) &   0.42 (2.02) &   0.00 (0.00) &   1.19 (3.48) &   0.04 (0.59) & 282.24 (105.72) \\ 
  $5 \times 10^{-7}$ &   0.00 (0.00) &   0.48 (3.06) &   0.00 (0.00) &   1.28 (5.10) &   0.00 (0.00) & 391.28 (159.65) \\ 
   \hline

   \multicolumn{7}{c}{$n=1000, R^2 = 0.15$}    \\ \hline
   
$10^{-3}$ &  0.24 (0.05) &  0.48 (0.07) &  0.42 (0.06) &  1.04 (0.10) &  0.46 (0.07) & 16.73 (2.26) \\ 
  $10^{-4}$ &  0.17 (0.13) &  0.47 (0.22) &  0.05 (0.07) &  1.06 (0.33) &  0.33 (0.18) & 49.76 (8.79) \\ 
  $10^{-5}$ &   0.11 (0.33) &   0.47 (0.68) &   0.00 (0.05) &   1.13 (1.06) &   0.21 (0.46) & 152.08 (33.59) \\ 
  $10^{-6}$ &   0.05 (0.71) &   0.47 (2.13) &   0.00 (0.00) &   1.24 (3.52) &   0.12 (1.09) & 474.34 (135.54) \\ 
  $5 \times 10^{-7}$ &   0.03 (0.77) &   0.53 (3.21) &   0.00 (0.00) &   1.23 (4.93) &   0.10 (1.41) & 669.21 (208.14) \\ 
   \hline

   \multicolumn{7}{c}{$n=1000, R^2 = 0.2$}    \\ \hline
   
$10^{-3}$ &  0.28 (0.05) &  0.51 (0.07) &  0.40 (0.06) &  1.04 (0.10) &  0.51 (0.07) & 19.59 (2.19) \\ 
  $10^{-4}$ &  0.22 (0.15) &  0.50 (0.22) &  0.04 (0.06) &  1.07 (0.34) &  0.39 (0.20) & 60.79 (8.92) \\ 
  $10^{-5}$ &   0.16 (0.39) &   0.50 (0.70) &   0.00 (0.03) &   1.12 (1.04) &   0.29 (0.54) & 193.25 (35.62) \\ 
  $10^{-6}$ &   0.12 (1.07) &   0.49 (2.17) &   0.00 (0.00) &   1.22 (3.43) &   0.23 (1.50) & 626.61 (146.57) \\ 
  $5 \times 10^{-7}$ &   0.09 (1.34) &   0.56 (3.30) &   0.00 (0.00) &   1.21 (4.89) &   0.18 (1.89) & 896.07 (228.65) \\ 
   \hline
\end{tabular}
}
\end{table}

\newpage

\newpage
\begin{table}[ht]
\footnotesize
\captionsetup[table]{labelsep=space, justification=raggedright, singlelinecheck=off}
\begin{threeparttable}

\caption{: Four mediation pathways identified by DACT after controlling the FWER at $0.05$. The exposure is adult SES and the outcome is HbA1c. The total number of mediators is 228,088. In the mediator-exposure model, we adjusted for age, sex, race and residual white blood cell types (neutrophils, B cells, T cells, and natural killer cells), and included the methylation chip and position as random effects to account for potential batch effects. In addition, we adjusted for the exposure in the outcome-mediator model. The estimated mediation effect is $\hat{\alpha}\hat{\beta}$ and the proportion of mediation effect is provided in the parenthesis. The 95\% confidence interval (CI) is calculated based on 1,000 bootstrap samples.}
\begin{tabular}{ccccccccc}
  \hline
CpG & Chr & Gene & \begin{tabular}[c]{@{}c@{}}UCSC RefGene\\ Group\end{tabular} & $\hat{\alpha}$ & $\hat{\beta}$& \begin{tabular}[c]{@{}c@{}}Mediation effect\\ (proportion)\end{tabular} & 95\% CI  & $p_{DACT}$\\ 
  \hline
cg10508317 & 17 & SOCS3 & Body  & -9.92E-02 & -1.78E-01 & 0.018 (0.18) & (0.009,0.029) & 5.35E-09 \\ 
  cg01288337 & 14 & RIN3 & Body & 5.55E-02 & 3.06E-01 & 0.017 (0.17)  & (0.009,0.028)  & 7.32E-09 \\ 
cg10244976 & 16 & LMF1 & Body  &  -1.19E-01 & -1.28E-01  & 0.015 (0.15)  & (0.007,0.025)  & 5.47E-08\\ 
 cg21263566 & 1 & TLCD4  & Body & 9.00E-02 & 1.59E-01 & 0.014 (0.14)   &  (0.006,0.024)  & 1.81E-07 \\ 
   \hline
\end{tabular}
\label{table_DACT}
\end{threeparttable}
\end{table}

\end{document}